\documentclass[journal]{IEEEtran}
\usepackage{graphicx}
\usepackage{booktabs}
\usepackage{subfigure}
\usepackage{amsmath}
\usepackage{algorithm}
\usepackage{algorithmic}
\usepackage{amssymb}
\usepackage{hyperref}
\hypersetup{
	colorlinks=true,
	linkcolor=blue,
	urlcolor=blue,
	citecolor=blue
}

\begin{document}
\title{Frequency-Compensated Network for Daily Arctic Sea Ice Concentration Prediction}
\author{
    Jialiang Zhang,
    Feng Gao, \emph{Member}, \emph{IEEE},
    Yanhai Gan,
    Junyu Dong, \emph{Member}, \emph{IEEE}, \\
    Qian Du, \emph{Fellow}, \emph{IEEE}

    \thanks{This work was supported in part by the National Science and Technology Major Project under Grant 2022ZD0117201, and in part by the Natural Science Foundation of Qingdao under Grant 23-2-1-222-ZYYD-JCH.

        Jialiang Zhang, Feng Gao, Yanhai Gan, and Junyu Dong are with the School of Computer Science and Technology, Ocean University of China, Qingdao 266100, China. (\textit{Corresponding author: Feng Gao})

        Qian Du is with the Department of Electrical and Computer Engineering, Mississippi State University, Starkville, MS 39762 USA.}}

\markboth{IEEE Transactions on Geoscience and Remote Sensing}
{Shell}

\maketitle

\begin{abstract}

    Accurately forecasting sea ice concentration (SIC) in the Arctic is critical to global ecosystem health and navigation safety. However, current methods still is confronted with two challenges: 1) these methods rarely explore the long-term feature dependencies in the frequency domain. 2) they can hardly preserve the high-frequency details, and the changes in the marginal area of the sea ice cannot be accurately captured. To this end, we present a Frequency-Compensated Network (FCNet) for Arctic SIC prediction on a daily basis. In particular, we design a dual-branch network, including branches for frequency feature extraction and convolutional feature extraction. For frequency feature extraction, we design an adaptive frequency filter block, which integrates trainable layers with Fourier-based filters. By adding frequency features, the FCNet can achieve refined prediction of edges and details. For convolutional feature extraction, we propose a high-frequency enhancement block to separate high and low-frequency information. Moreover, high-frequency features are enhanced via channel-wise attention, and temporal attention unit is employed for low-frequency feature extraction to capture long-range sea ice changes. Extensive experiments are conducted on a satellite-derived daily SIC dataset, and the results verify the effectiveness of the proposed FCNet. Our codes and data will be made public available at: \url{https://github.com/oucailab/FCNet}.

\end{abstract}


\IEEEpeerreviewmaketitle

\section{Introduction}

\IEEEPARstart{T}HE Arctic sea ice is an important component of the global climate system \cite{walsh2020extreme} \cite{in14tgrs}, since its contributions to the Earth's reflectivity and potential influence on global weather patterns. As the Arctic sea ice extent decreases, the reflection of sunlight diminishes, leading to increased oceanic heat absorption and exacerbating global warming \cite{vihma2014effects} \cite{dyh22tgrs}. The influx of freshwater from melting ice contributes to sea level rise, and the reduction in sea ice diminishes habitats for Arctic fauna, adversely impacting the entire ecological chain \cite{lannuzel2020future} \cite{hyl22jars}. Meanwhile, the melting sea ice in the Arctic is opening up new shipping lanes and altering global shipping routes with the benefits of reducing distances for certain pathways \cite{yang2019improving} \cite{wj22jars}. It is expected that by 2030, the navigation period of the Arctic waterway is to be extended from the current four months to more than half a year \cite{chen22anth}. Therefore, sea ice prediction in the Arctic, especially short-term sea ice concentration, is of great significance for the ecosystem health and navigation safety \cite{bhb18jstars} \cite{cl19tgrs} \cite{bk21tgrs}.

\begin{figure}
    \centering
    \includegraphics[width=\linewidth]{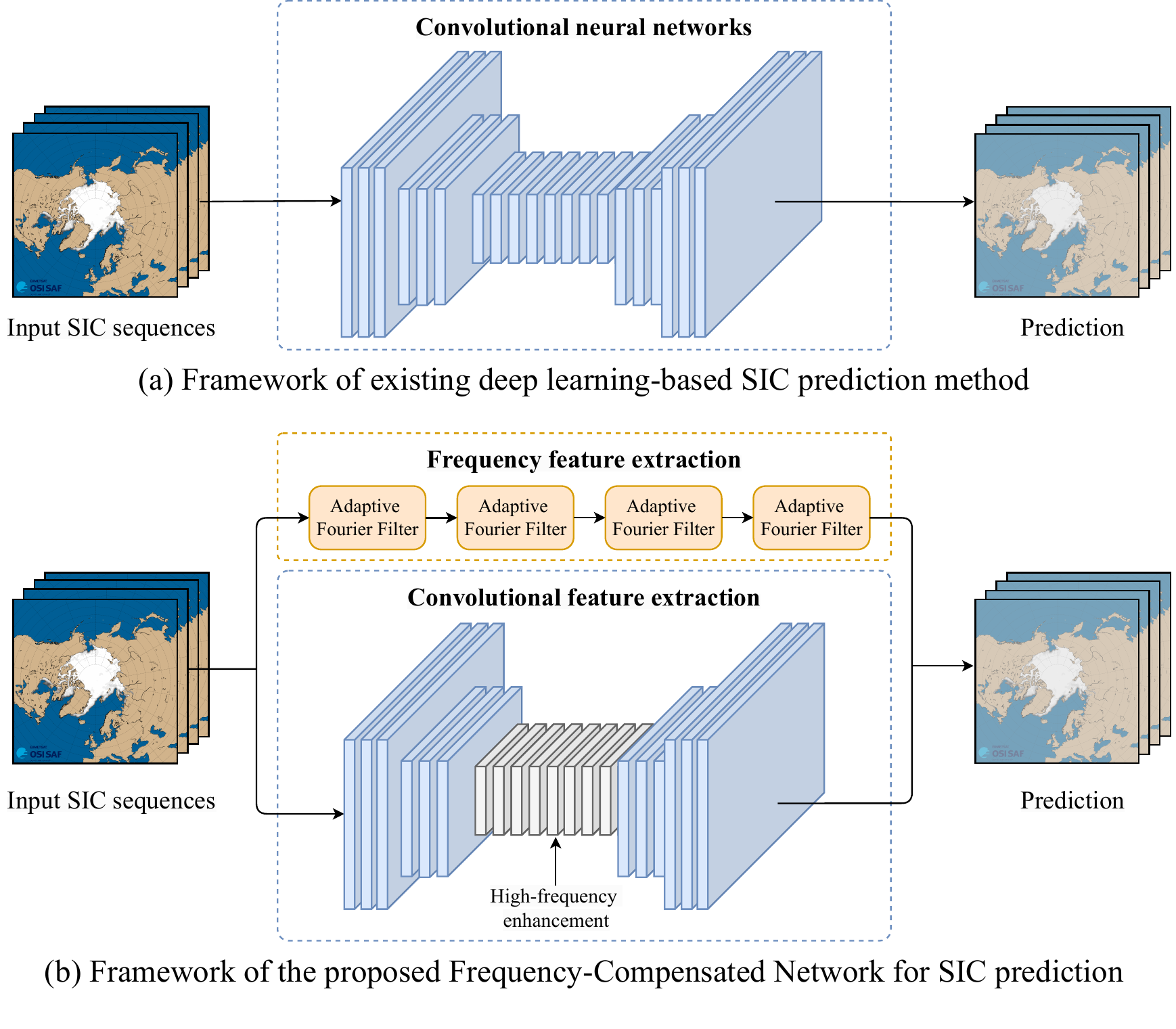}
    \caption{Comparisons of the proposed Frequency-Compensated Network (FCNet) with existing deep learning-based SIC prediction methods. }
    \label{FCNet和传统方法对比}
\end{figure}

Over the past decades, many efforts have been dedicated to solving the problem of SIC prediction. Existing SIC prediction models can be roughly divided into three categories, i.e., numerical models, statistical models, and  machine/deep learning models \cite{yuan16jc}. Numerical models are based on physical links between temperature changes, humidity transport, wind field, ocean heat fluxes, and sea ice dynamics, which together influence the prediction of Arctic sea ice concentration. They are the mainstream methods for Arctic sea ice prediction at the synoptic scale \cite{her15jgr}. Statistical models make long-term predictions by building the relationships between oceanic, atmospheric, and sea ice parameters based on historical data. Most statistical models are linear models, which can hardly capture the nonlinear relationship between the historical data and predicted results \cite{wang16jc}.

Deep learning has emerged as a powerful tool in oceanography and remote sensing, aiding in the comprehension of global climate system and climate change \cite{mengyx23tnnls} \cite{mengyx23tgrs}. For some regional sea ice and climate mechanisms that humans have not yet mastered, classical numerical or statistical models can hardly predict accurately. Deep learning models can effectively handle complex, high-dimensional data and learn intricate patterns within the data \cite{phd23jstars} \cite{zhp23grsl}. They have been employed in sea ice detection \cite{gao2019sea}, classification \cite{zhang24jstars}, and prediction \cite{kim2020pred}. In particular, ConvLSTM \cite{li21cl} \cite{he22oceans}, Convolutional Neural Networks (CNNs) \cite{ren21igarss} and Attention Mechanism \cite{li24tgrs} have been employed for Arctic SIC prediction. These models have achieved promising results for SIC prediction. Therefore, in this paper, we are dedicated to develop robust deep learning-based SIC prediction model.

Indeed, existing deep learning-based models have demonstrate remarkable success in Arctic sea ice prediction. However, it is non-trivial to design a robust SIC prediction model due to the following challenges: 1) \textbf{Frequency domain features are rarely explored.}  Existing methods commonly learn feature interactions among spatial-temporal locations from the raw data. Although notable efforts have been dedicated to SIC prediction via ConvLSTM and CNNs \cite{li21cl} \cite{ren21igarss} \cite{li24tgrs}, these methods suffer from weak flexibility and low efficiency since long-term dependencies in the frequency domain are not explored. 2) \textbf{Lacking in the enhancement of high-frequency details.} Existing methods usually focus on global regularity of the SIC prediction results. However, these methods fail to preserve the high-frequency details, and the changes in the marginal area of sea ice can hardly been captured.

To address the above issues, we propose a Frequency-Compensated Network (FCNet) for daily Arctic SIC prediction. As shown in Fig. \ref{FCNet和传统方法对比}, the proposed FCNet comprises two feature extraction branches, including one for frequency feature extraction and the other for convolutional feature extraction. In the frequency feature extraction branch, we explore the long-range dependencies in the frequency domain. We design an Adaptive Frequency Filter Block (AFFB), which integrates trainable neural networks with Fourier-based filters. In addition, we design a frequency loss to enhance the focus on changes in the marginal area of sea ice. In the convolutional feature extraction branch, an encoder-decoder network with skip connections is employed. We design High-Frequency Enhancement Blocks (HFEB) to enhance the high-frequency feature information. We perform extensive experiments on OSI-450-a dataset, where the proposed FCNet achieves superior performance over existing methods, demonstrating the effectiveness of the using of frequency features.

To summarize, our contributions are two-folded:

\begin{itemize}

    \item To the best of our knowledge, FCNet is the first work to investigate the frequency domain features for daily SIC prediction. By using adaptive Fourier filters, FCNet achieves refined prediction of edges and details.

    \item We propose HFEB, which  use pooling to separate high and low-frequency information. High-frequency features are enhanced via channel-wise attention. Temporal attention unit is employed for low-frequency feature extraction to capture long-range sea ice changes.


\end{itemize}

\section{Related Works}

\subsection{Deep Learning-Based Sea Ice Concentration Prediction}

Deep learning models have been rapidly developed as a powerful tool to capture the nonlinear relationships between data and correlated predictions. Recently, deep learning-based methods have been successfully applied to sea ice forecasting in both the Arctic and Antarctic \cite{kim2020pred}. Andersson et al. \cite{andersson2021seasonal} proposed a deep learning sea ice prediction system (IceNet) that predicted the next six months of sea ice concentration maps, using U-Net for feature learning, and outperformed the ECMWF SEAS5 dynamic seasonal prediction system. Grigoryev et al. \cite{gri22rs} developed a short-term sea ice prediction method based on U-Net, incorporating sea ice observations and weather forecasts. Asadi et al. \cite{tc-16-3753-2022} used sequence-to-sequence learning and ERA5 data to predict seasonal sea ice presence in the Hudson Bay region, providing daily spatial maps of sea ice presence probability for lead times up to 90 days, demonstrating the ability to predict sea ice presence probabilities with skill during the break-up season. Ren et al. \cite{9780401} presented an encoder-decoder network with a spatio-temporal attention module for SIC prediction that improves feature representations of extracted spatio-temporal dependencies. Yuan et al. \cite{yuan24om} integrated ConvLSTM into U-Net to improve the accuracy of SIC prediction. Ren et al. \cite{ren23tgrs} proposed a physically constrained loss function to optimize the details of the SIC prediction results, effectively improving the visual details. Li et al. \cite{li24tgrs} presented a SIC prediction network by integrating the convolutional block attention module and the temporal feature convolution module.

\subsection{Frequency Information Analysis}

Frequency domain information is critical for image analysis \cite{yys24tgrs} \cite{ali24spl}. Fourier analysis highlights a phenomenon known as spectral bias \cite{rahaman2019spectral} \cite{mildenhall2021nerf}, which is a neural network's tendency to learn low-frequency functions more efficiently. Using coordinate-based MLPs, Fourier features \cite{rahimi2007random} and positional encoding \cite{mildenhall2021nerf} \cite{vaswani2017attention} have been employed to recover missing high frequencies in image regression tasks.

Additionally, some studies combine frequency analysis with network compression \cite{gueguen2018faster} \cite{xu2020learning} \cite{yoo2019photorealistic} and feature reduction \cite{levinskis2013convolutional} to accelerate network training and inference. The applications of frequency domain analysis have further expanded to include media forensics, super-resolution, generalization analysis, magnetic resonance imaging, and image scaling. Despite extensive exploration of frequency information in vision tasks, the investigation of improving sequential data prediction via the frequency domain features is still limited.

\begin{figure*}[htbp]
    \centering
    \includegraphics [width=6.4in]{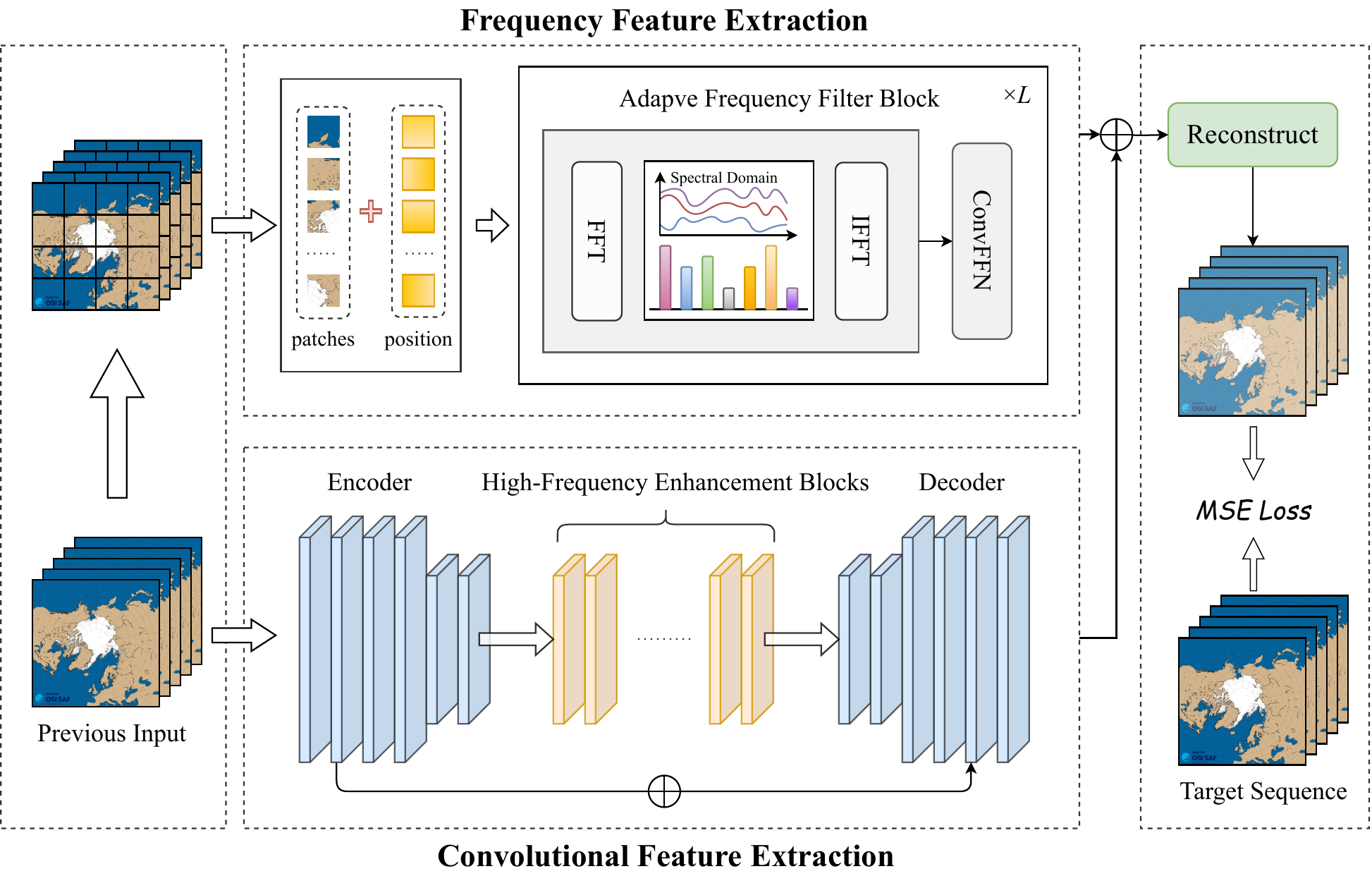}
    \caption{The framework overview of the Frequency-Compensated Network (FCNet) for daily Arctic SIC prediction. The proposed FCNet comprises three components: the frequency extraction branch, the convolutional feature extraction branch, and the refinement block.}
    \label{FCNet架构总览}
\end{figure*}

\section{Methodology}

\subsection{Problem Definition}

In this paper, short-term Arctic SIC prediction aims to infer the future sea ice daily data using the previous ones. Concretely, given a specific $t$-th day and a sequence of daily Arctic sea ice data $\mathbf{X}_{t,T}=\left \{ x_i \right \}^{t}_{t-T+1}$ for the past $T$ days, our goal is to predict a sequence of sea ice data $\mathbf{Y}_{t,{T}'}=\left \{ x_i \right \}^{t+{T}'}_{t+1}$ for the subsequent ${T}'$ days starting from ($t+1$)-th day. Here, ${x_i}\in \mathbb{R}^{C\times H\times W}$ denotes the $i$-th three dimensional (3D) SIC data with variable $C$, height $H$ and width $W$. In this paper, the $C$ dimension contains only one variable, sea ice concentration. A four dimensional (4D) sea ice sequence data is represented as a tensor, i.e., $\mathbf{X}_{t,T} \in \mathbb{R}^{T\times C\times H\times W} $ and $\mathbf{Y}_{t,{T}'} \in \mathbb{R}^{{T}'\times C\times H\times W}$.

A model with learnable parameters $\Theta$ learns the mapping $\mathcal{F}_{\Theta}:\mathbf{X}_{t, T}\mapsto \mathbf{Y}_{t, {T}'}$ by investigating spatial and temporal dependencies. In our scenario, the mapping $\mathcal{F}_{\Theta}$ is a neural network model that is trained to minimize the discrepancy between the predicted SIC data and the ground truth. The optimal parameters ${\Theta}^*$ can be derived through the following equation:

\begin{equation}
    \Theta^{*}=\arg \min _{\Theta} \mathcal{L}\left(\mathcal{F}_{\Theta}\left(\mathbf{X}_{t, T}\right), \mathbf{Y}_{t, T^{\prime}}\right)
\end{equation}
where $\mathcal{L}$ is a loss function used to assess differences between the predicted SIC data and the ground truth.

\subsection{Overview of the proposed FCNet}

The key to predicting future sea ice concentration series lies in designing a model that can capture spatial-temporal dependencies from historical SIC series. The model can then use the captured spatial-temporal dependence information to predict future SIC states. As shown in Fig. \ref{FCNet架构总览}, the proposed FCNet  comprises three primary components: the frequency feature extraction branch, the convolutional feature extraction branch, and the refinement block.

\textbf{Frequency Feature Extraction Branch.} The frequency feature extraction branch is focused on extraction features in the Fourier domain. Specifically, given the SIC sequence $\mathbf{X}_{t,T} \in \mathbb{R}^{T\times C\times H\times W}$, we first divide the daily sea ice data in the sequence into several non-overlapping patches of a fixed size $h\times w$, and linearly project these patches into patch embeddings of $\mathbf{E}^{pat} \in \mathbb{R}^{T\times h \times w\times d}$, where $d$ denotes the dimension of the embeddings. The position embeddings $\mathbf{E}^{pos} $ with the same size of $\mathbf{E}^{pat}$ are applied to generate an initial token representation E as follows:

\begin{equation}
    \mathbf{E} = \mathbf{E}^{pat} + \mathbf{E}^{pos}
\end{equation}
Then, the token representations are fed into the Adaptive Frequency Filter Blocks (AFFB) for frequency feature extraction. The AFFB is repeated $L$ times to exploit deep frequency features. AFFB is the key component in the frequency feature extraction branch, and will be introduced in detail in Section \ref{Section:AFFB}.

\textbf{Convolutional Feature Extraction Branch.} The convolutional feature extraction branch extracts features from the raw data through convolutions. It is an encoder-decoder network with skip connections. The convolution feature extraction branch consists of encoder, high-frequency enhancement blocks (HFEB), and decoder. The encoder utilizes four  ConvNormSiLU blocks to  extraction spatial features. Given the SIC sequence $\mathbf{Z}_{i-1}$, the computation can be expressed as:
\begin{equation}
    \mathbf{Z}_{i}=\sigma ( \text{GroupNorm}  ( \text{Conv} ( \mathbf{Z}_{i-1} )))
\end{equation}
where  $\mathbf{Z}_{i-1}$ and $\mathbf{Z}_i$ denote the input and output of the $i$-th block ($1\leq i \leq 4$), respectively, and $\sigma$ is the SiLU activation. The decoder contains four ConvNormSiLU blocks for feature reconstruction. Different from the encoder, Conv2d is replaced by ConvTranspose2d. HFEB is the critical component in the convolutional feature extraction branch, and will be detailed in Section \ref{Section:HFEB}.

\textbf{Refinement Block and Loss Function.} In the refinement block, two $3\times 3$ depth-wise convolutional layers are employed to refine the spatial details of the SIC prediction results. To further improve the spatial details of the prediction results, we design a frequency loss to adaptively adjust  the weights of different frequency components. Details of the loss function will be detailed in Section \ref{Section:Loss Function}.

\begin{figure}[htbp]
    \centering
    \includegraphics [width=\linewidth]{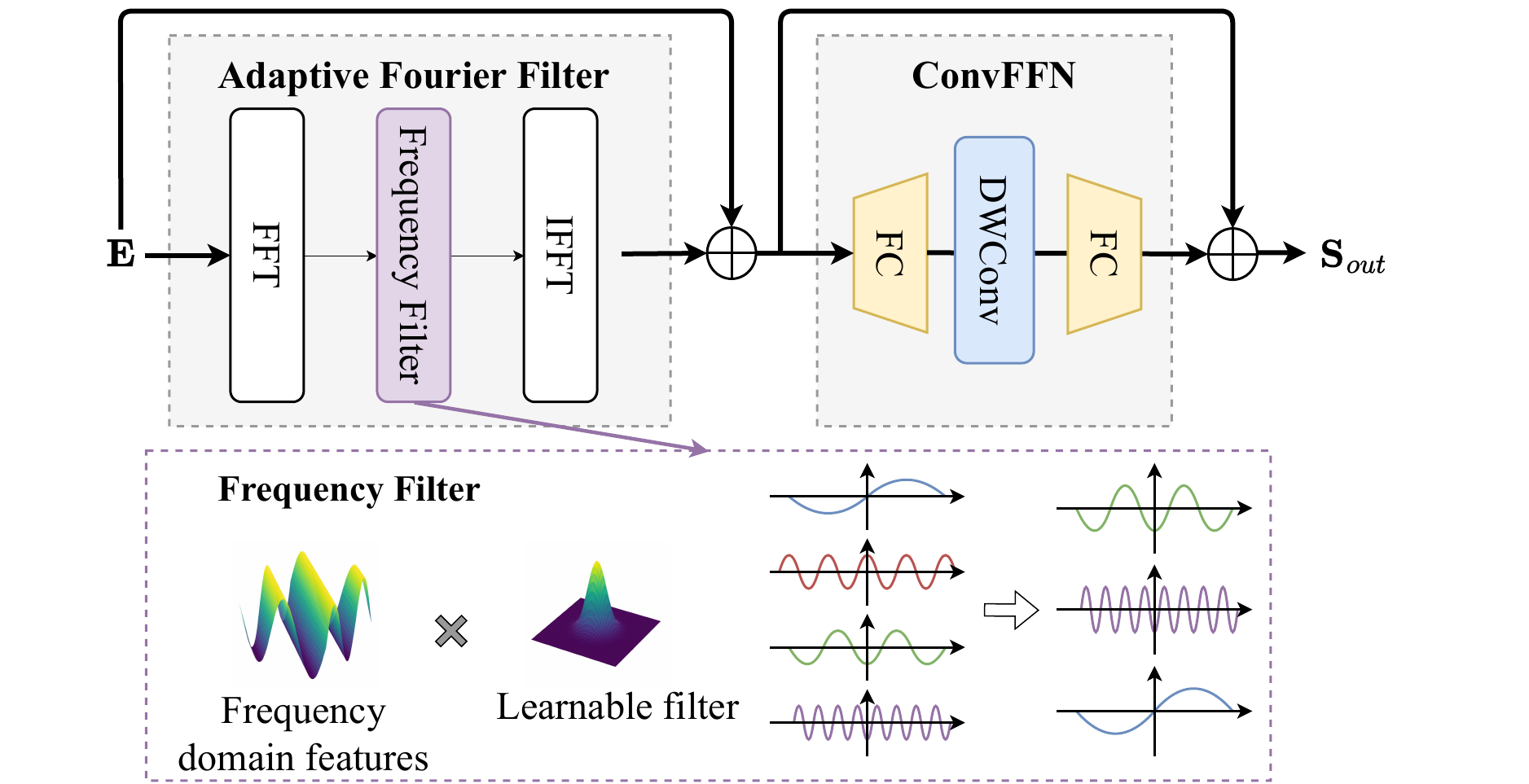}
    \caption{Illustration of the Adaptive Frequency Filter Blocks (AFFB).}
    \label{AFFB}
\end{figure}

\subsection{Adaptive Frequency Filter Blocks (AFFB)}
\label{Section:AFFB}

Our primary objective is to leverage long-range dependencies in the frequency domain for SIC prediction. Inspired by \cite{li20fno} and \cite{guibas21ada}, we integrate trainable neural networks with Fourier-based filters. As shown in Fig. \ref{AFFB}, the AFFB mainly consists of two parts: Adaptive Fourier Filter and ConvFFN.

\textbf{Adaptive Fourier Filter.} We first employ the Fast Fourier Transform (FFT) to the input feature $\mathbf{E}$, and converts the input signals from the spatial domain to the frequency domain. The frequency domain feature $f(u,v)$ is computed as follows:
\begin{equation}
    f(u, v)= \sum_{x=0}^{H-1} \sum_{y=0}^{W-1} \mathbf{E}(x, y) e^{-j 2 \pi(\frac{ux}{H}+\frac{vy}{W})}
\end{equation}
where $ \mathbf{E}(x, y) $ represents the input feature map at spatial coordinates $ (x,y) $, where $ x $ and $ y $ are the spatial indices in the height and width dimensions, respectively. $ f(u,v) $: This denotes the frequency domain feature at frequency coordinates $ (u,v) $, where $ u $ and $ v $ are the frequency indices in the horizontal and vertical directions, respectively. $ j $ is the imaginary unit.

In the frequency domain, data is represented as fluctuations of different frequencies, providing an opportunity to identify periodic patterns. It has been demonstrated that high-frequency components are typically associated with edges, textures, and fine details in images \cite{bai22eccv} \cite{yin19nips}. Therefore, to extract the high-frequency portions most relevant to the temporal features of sea ice, we filter the frequency domain features with the aim of reducing noise or eliminating irrelevant frequency components as follows:

\begin{equation}
    \hat{f} \left(u, v\right ) = \mathcal{W} \odot f\left(u, v\right )
\end{equation}
where $\mathcal{W}$ is the weight matrix. It is initialized randomly, and has the same dimensions as the filter $f(u, v)$. Here, $\odot$ denotes the element-wise multiplication.

Subsequently, we perform an Inverse Fast Fourier Transform (IFFT) to convert the features from the frequency domain to the spatial domain as follows:
\begin{equation}
    \mathbf{S}(x, y) =\frac{1}{HW} \sum_{u=0}^{H-1} \sum_{v=0}^{W-1} \hat{f}(u, v) e^{j 2 \pi(\frac{ux}{H}+\frac{vy}{W})}
\end{equation}

\textbf{ConvFFN.} To better leverage the latent spatial-temporal information to reconstruct the prediction results, we introduce ConvFFN to perform non-linear feature transformation. Inspired by PVT \cite{pvtv2}, the ConvFFN consists of two fully connected layers, and one $3\times3$ depth-wise convolution, as shown in Fig. \ref{AFFB}. The computation of ConvFFN is as follows:
\begin{equation}
    \mathbf{S}_{out}=\textrm{FC}(\textrm{DWConv} ( \textrm{FC}( \mathbf{S} ))) +\mathbf{S}
\end{equation}
where \textrm{FC} denotes fully connected layer and \textrm{DWConv} refers to depth separable convolutional.

\begin{figure}[]
    \centering
    \includegraphics[width=\linewidth]{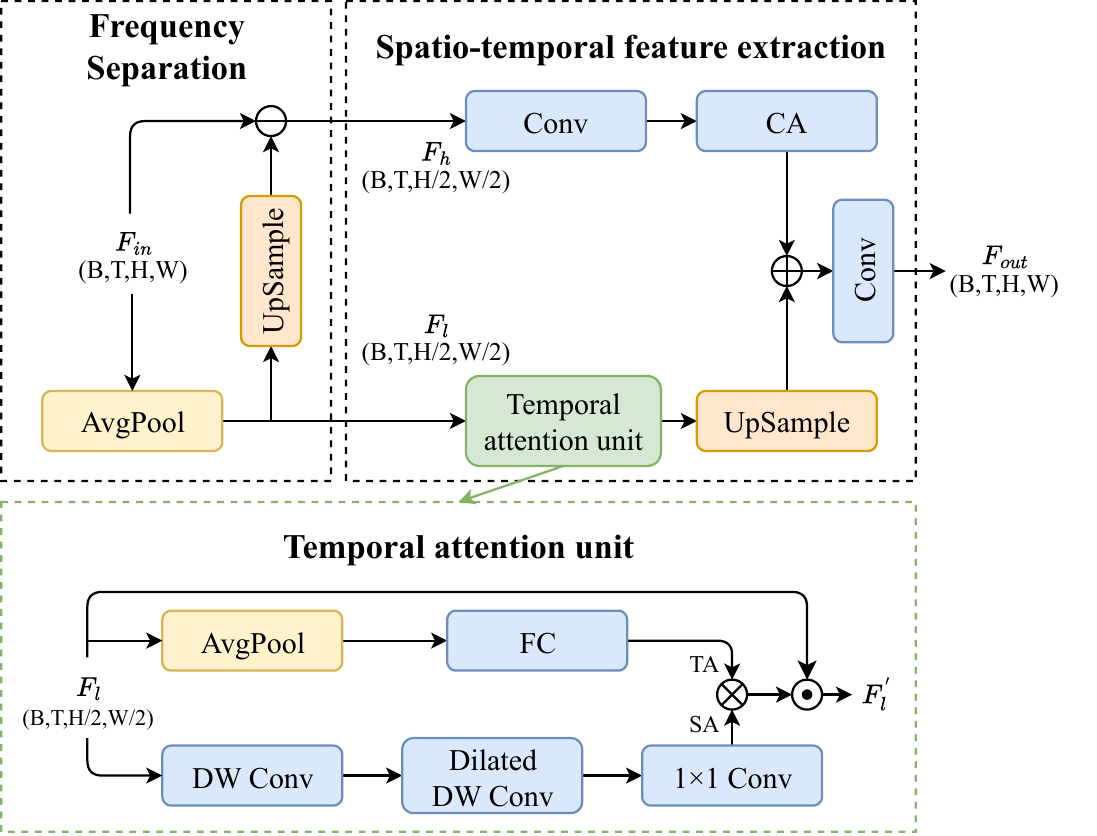}
    \caption{Details of the proposed High-Frequency Enhancement Block (HFEB).}
    \label{HFEB}
\end{figure}

\subsection{High-Frequency Enhancement Block (HFEB)}
\label{Section:HFEB}

In SIC prediction, enhancing high-frequency feature information is critical. In the proposed HFEB, we use pooling for high-frequency and low-frequency feature separation. As shown in Fig. \ref{HFEB}, HFEB includes two components: frequency separation, spatial-temporal feature extraction.

\textbf{Frequency Separation.} In the proposed HFEB, we employ pooling for high-frequency and low-frequency feature separation. Specifically, a pooling layer down-samples the input feature with dimensions $(B,T,H,W)$ to obtain the low-frequency feature $F_l$ at a reduced resolution $(B, T, \frac{H}{2}, \frac{W}{2})$.

The high-frequency feature $F_h$ is subsequently computed by subtracting $F_l$ from the original features $F$. Therefore, we captures the data's high and low-frequency characteristics, as follows:
\begin{equation}
    F_l=\textrm{Pool}(F)
\end{equation}
\begin{equation}
    F_h=F-\textrm{Upsample}(F_l)
\end{equation}
where Pool denotes the down-sampling operation using average pooling, and Upsample refers to the up-sampling operation using bilinear interpolation.

\textbf{Spatial-Temporal Feature Extraction.} After obtaining the high-frequency feature $F_h$, using channel-wise attention to fully exploit the high-frequency information and get $F'_h$.

For the low-frequency feature $F_l$, we use the temporal attention unit \cite{tan23cvpr} to capture long-range sea ice changes by expanding the receptive field. Concretely, small kernel depth-wise convolution (DWConv), depth-wise convolution with dilations (DWDConv), and point-wise convolution to are used to model the large kernel convolution. Furthermore, the dynamic attention is employed to learn the temporal evolution along the timeline. The detailed computation for low-frequency feature is as follows:
\begin{equation}
    \text{SA} = \text{Conv}_{1 \times 1}(\text{DWDConv}(\text{DWConv}(F_l)))
\end{equation}

\begin{equation}
    \text{TA}=\text{FC}(\text{AvgPool}(F_l))
\end{equation}

\begin{equation}
    F'_l =(\text{SA} \otimes \text{TA}) \odot F_l
\end{equation}
where SA denotes the large kernel attention, TA denotes the temporal attention, FC is the fully connected layer, AvgPool is the average pooling, $\otimes$ is the matrix multiplication, and $\odot$ is the element-wise multiplication.

Finally, the high-frequency feature $F'_h$ and low-frequency feature $F'_l$ are combined by element-wise summation to generate the output feature $F_{out}$ as follows:

\begin{equation}
    F_{out} = \text{Conv}_{1\times1}(F'_h + \textrm{Upsample}(F'_l))
\end{equation}
where $\textrm{Conv}_{1\times1}$ denotes the point-wise convolution.

\subsection{Loss Function}
\label{Section:Loss Function}

In this paper, we investigate the frequency domain gap between the ground truth SIC and the predicted SIC to optimize the frequency feature extraction branch. Due to the  emphasis on a phenomenon known as spectral bias through Fourier analysis \cite{rahaman2019spectral} \cite{mildenhall2021nerf}, which is a learning bias of neural networks towards low-frequency functions, the prediction models tend to avoid difficult frequency components and converge to suboptimal results. To address this problem, we compute the frequency distance between the ground truth SIC and the predicted SIC as follows:

\begin{equation}
    d(\textrm{F}_g, \textrm{F}_p) = \left | \textrm{F}_g(u,v)- \textrm{F}_p(u,v)\right | ^{2}
\end{equation}
where $\textrm{F}_g$ is the ground truth SIC in the frequency domain, and $\textrm{F}_p$ is the output of the frequency feature extraction branch in the frequency domain.

Our proposed frequency distance metric provides a quantitative measure to compare the ground truth SIC and predicted SIC in the frequency domain. However, employing frequency distance directly as a loss function results in model bias towards easily predictable frequencies, as this approach assigns equal weight to all frequency components. To address this limitation and enable adaptive adjustment of weights across different frequency components, we introduce a spectral weight matrix. This matrix is dynamically determined by the non-uniform distribution of training losses across frequencies during the learning process. The spectral weight matrix maintains the same dimensionality as the frequency spectrum. We define the weight of the spatial frequency at point $(u, v)$ as:
\begin{equation}
    w(u,v) = \left | \log \left ( \left |\text{F}_g(u,v)- \text{F}_p(u,v)\right |\right )\right |
\end{equation}
To ensure consistent scaling across frequencies, we normalize the matrix values to the range [0,1]. In this normalized scheme, a weight of 1 corresponds to the frequency component with the maximum current loss, while frequencies that are more easily predicted receive proportionally lower weights. Notably, we prevent gradient flow through the spectral weights matrix by implementing gradient locking, thereby restricting its role to frequency-specific weighting without affecting the back-propagation dynamics.

By taking the Hadamard product of the spectral weight matrix and the frequency distance matrix, we compute the frequency loss as:
\begin{equation}
    \mathcal{L}_{freq} = \frac{1}{HW}\sum_{u=0}^{H-1}\sum_{v=0}^{W-1} w(u,v) || \text{F}_g(u,v)- \text{F}_p(u,v) || ^2
\end{equation}
where the feature size is $H \times W$. The frequency loss can be considered as the weighted average of the frequency distance between the ground truth SIC and the predicted SIC.

Beside the frequency loss, we use the MSE loss as the prediction loss $\mathcal{L}_{pred}$ to minimize the error between the ground truth SIC and the predicted SIC.

The overall loss function for our proposed FCNet is defined as:
\begin{equation}
    \mathcal{L} = \mathcal{L}_{pred}+\lambda \mathcal{L}_{freq}
\end{equation}
where $\lambda$ is a weight parameter to balance the prediction loss and frequency loss.

To summarize the workflow of the proposed FCNet, we present the corresponding pseudocode in Algorithm \ref{alg:FCNet}. The algorithm takes a continuous sequence of real sea ice concentration $\mathbf{X}_{t,T}$ as input and outputs the predicted sea ice concentration sequence $\mathbf{Y}_{t, {T}'}$. The network first extracts frequency features using Fourier transform and adaptive filtering, followed by convolutional feature extraction for spatial-temporal learning. Finally, the extracted features are fused to generate the final prediction.

\begin{algorithm}
    \caption{FCNet: Frequency-Compensated Network}
    \label{alg:FCNet}
    \renewcommand{\algorithmicrequire}{\textbf{Input:}}
    \renewcommand{\algorithmicensure}{\textbf{Output:}}
    \begin{algorithmic}[1]
        \REQUIRE Sequence of real sea ice concentration $\mathbf{X}_{t,T}$
        \ENSURE Sequence of predicted sea ice concentration $\mathbf{Y}^{t, {T}'}$

        \textbf{Frequency Feature Extraction:}
        \STATE $f(u, v) \leftarrow \text{FFT}(\mathbf{X}_{t,T})$
        \STATE $f_{\text{filtered}} \leftarrow f(u, v) \cdot W_{learnable}$
        \STATE $S(x, y) \leftarrow \text{IFFT}(f_{\text{filtered}})$
        \STATE $S_{out} \leftarrow \textrm{FC}(\textrm{DWConv} ( \textrm{FC}( S(x, y) ))) +S(x, y)$

        \textbf{Convolutional Feature Extraction:}
        \STATE $F \leftarrow \text{CNNEncoder}(\mathbf{X}_{t,T})$
        \STATE $F_l \leftarrow \text{AvgPool}(F)$
        \STATE $F_h \leftarrow F - \text{Upsample}(F_l)$

        \STATE $TA \leftarrow \text{TemporalAttention}(F_l)$
        \STATE $SA \leftarrow \text{SpatialAttention}(F_l)$
        \STATE $F'_l =(\text{SA} \otimes \text{TA}) \odot F_l$
        \STATE $F'_h \leftarrow \text{CA}(F_h)$
        \STATE $F_{\text{combined}} \leftarrow F_h' + \text{Upsample}(F_l')$
        \STATE $F_{out} \leftarrow \text{CNNDecoder}(F_{\text{combined}})$

        \textbf{Feature Fusion and Prediction:}
        \STATE $\mathbf{Y}^{t, {T}'} \leftarrow F_{out} + S_{out}$

        \RETURN $\mathbf{Y}^{t, {T}'}$
    \end{algorithmic}
\end{algorithm}

\section{Experimental Results and Analysis}

\subsection{Experiment Settings}

\begin{figure}[htbp]
    \centering
    \includegraphics [width=3.2in]{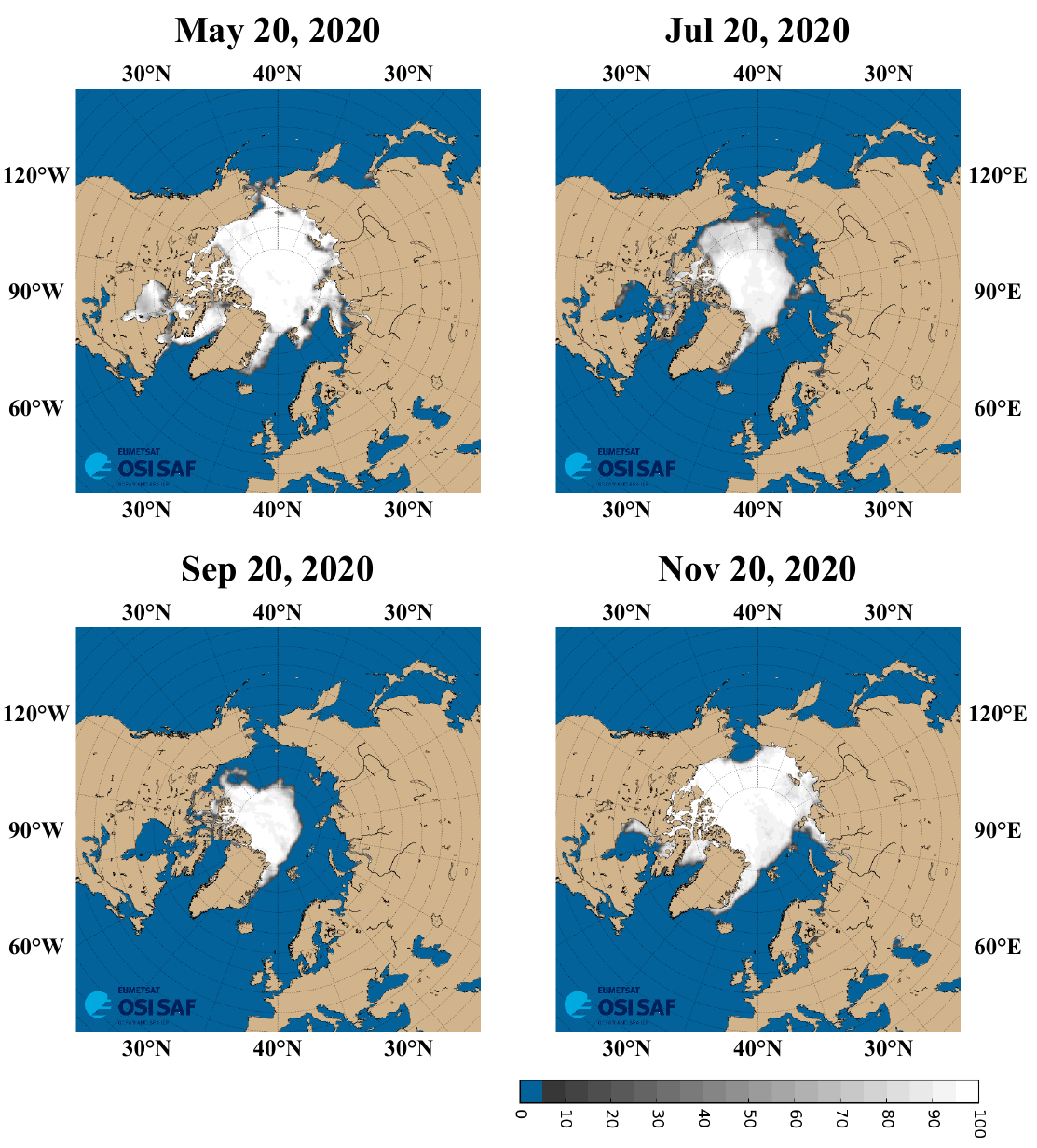}
    \caption{Daily sea ice concentration data from the OSI-450-a dataset.}
    \label{数据集样例}
\end{figure}

\textbf{Dataset Description.} In this paper, we use the OSI-450-a \cite{OSI} dataset from the OSI SAF Global Sea Ice Concentration Climate Data Records. This dataset provides global SIC information at a resolution of 25 km from 1978 to 2020, leveraging coarse resolution imagery from SMMR, SSM/I, and SSMIS satellites, supplemented by the latest ERA5 reanalysis data. OSI-450-a is a fixed-length climate data record that amalgamates recalibrated satellite data with ERA5 reanalysis to provide extensive and long-term observations of global sea ice variations. We employ this dataset to scrutinize the spatial-temporal changes in SIC data. Fig. \ref{数据集样例} illustrates examples of SIC data in 2020.

\textbf{AMAP Area.} To ensure comprehensive coverage of the entire sea ice extent, we conducted model evaluations within the specified Arctic Monitoring and Assessment Programme (AMAP) area \cite{amap}. The AMAP area encompasses land and ocean regions north of the Arctic Circle (66°32'N), north of 62°N latitude in Asia, and north of 60°N latitude in North America. This dataset is pivotal for our research as it provides a robust foundation for analyzing the historical trends and patterns of SIC data, which is essential for developing and validating our FCNet. Using the AMAP area in our experiments ensures that our models are thoroughly tested against a diverse and representative set of environmental conditions, thereby confirming the scalability and reliability of our predictions.

\textbf{Evaluation Metrics.} To evaluate the performance of our FCNet, we have chosen three metrics: Mean Absolute Error (MAE), Root Mean Square Error (RMSE), and Nash-Sutcliffe Efficiency (NSE). MAE quantifies the average absolute deviation between the predicted values and the actual observations, providing a critical measure of sea ice predictive accuracy. RMSE gauges the average squared difference between the predicted and ground truth values, offering a more sensitive metric for predictive errors and is commonly used to evaluate model accuracy. NSE is a widely adopted metric to assess the predictive model's fidelity to observations. MAE reflects the average magnitude of predictive errors, with lower values indicating fewer predictive errors and greater predictive accuracy. Smaller RMSE values signify higher predictive accuracy. NSE takes into account both bias and variance between predicted and observed values, with values approaching one indicating excellent predictive performance, while negative values suggest performance inferior to that of a simple mean prediction. In this paper, MAE is calculated as the spatial average over the AMAP area grid, followed by temporal averaging. We denote $\text{SIC}_p$ as the predicted values and $\text{SIC}_g$ as the ground truth values, and use these to compute MAE, RMSE, and NSE metrics.

\begin{figure*}[htbp]
    \centering
    \includegraphics[width=6.4in]{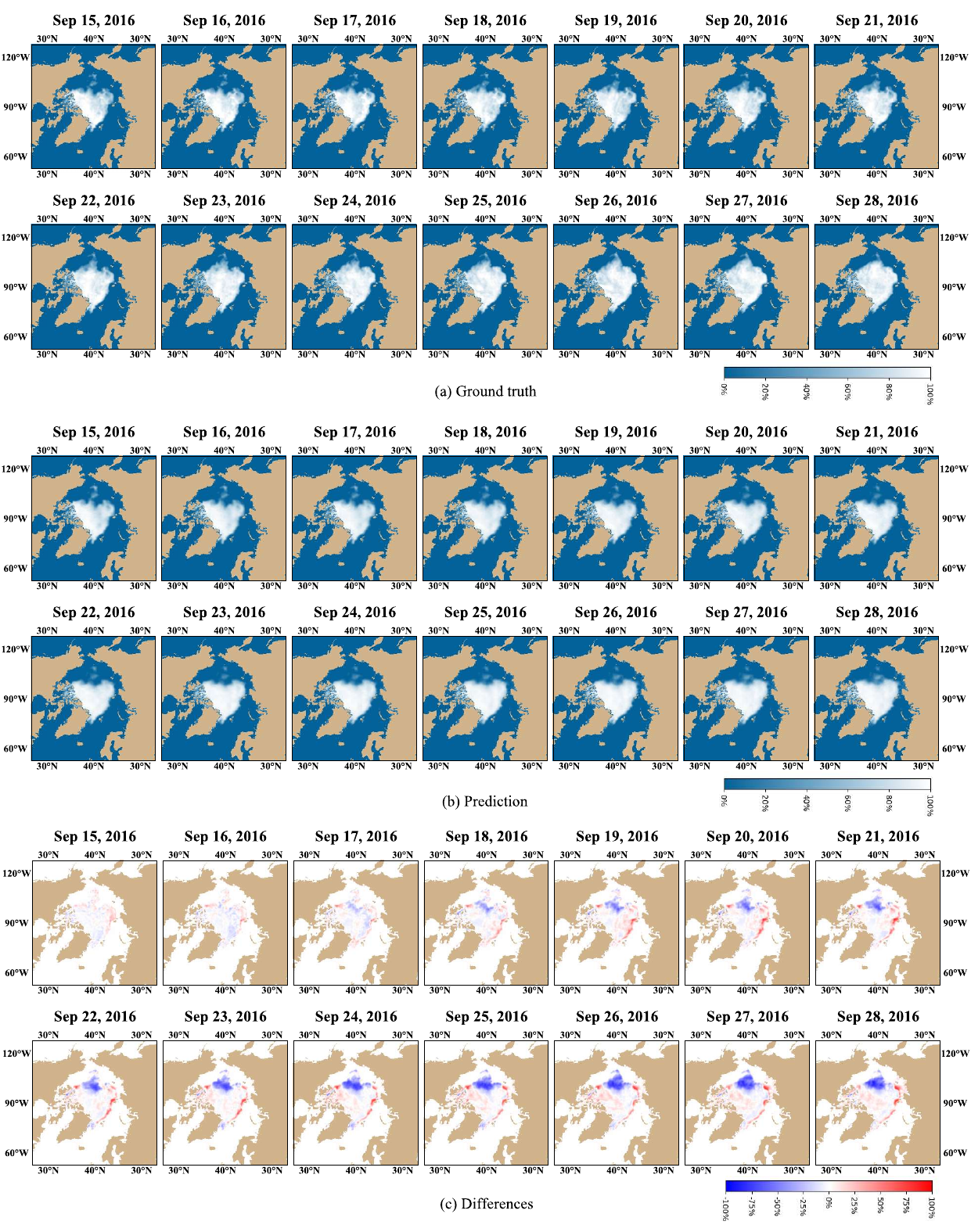}
    \caption{Details of the SIC prediction results for September 15–28, 2016. (a) Ground Truth. (b) Predictions. (c) Differences.}
    \label{fig_vis2016}
\end{figure*}

\begin{figure*}[htbp]
    \centering
    \includegraphics[width=6.4in]{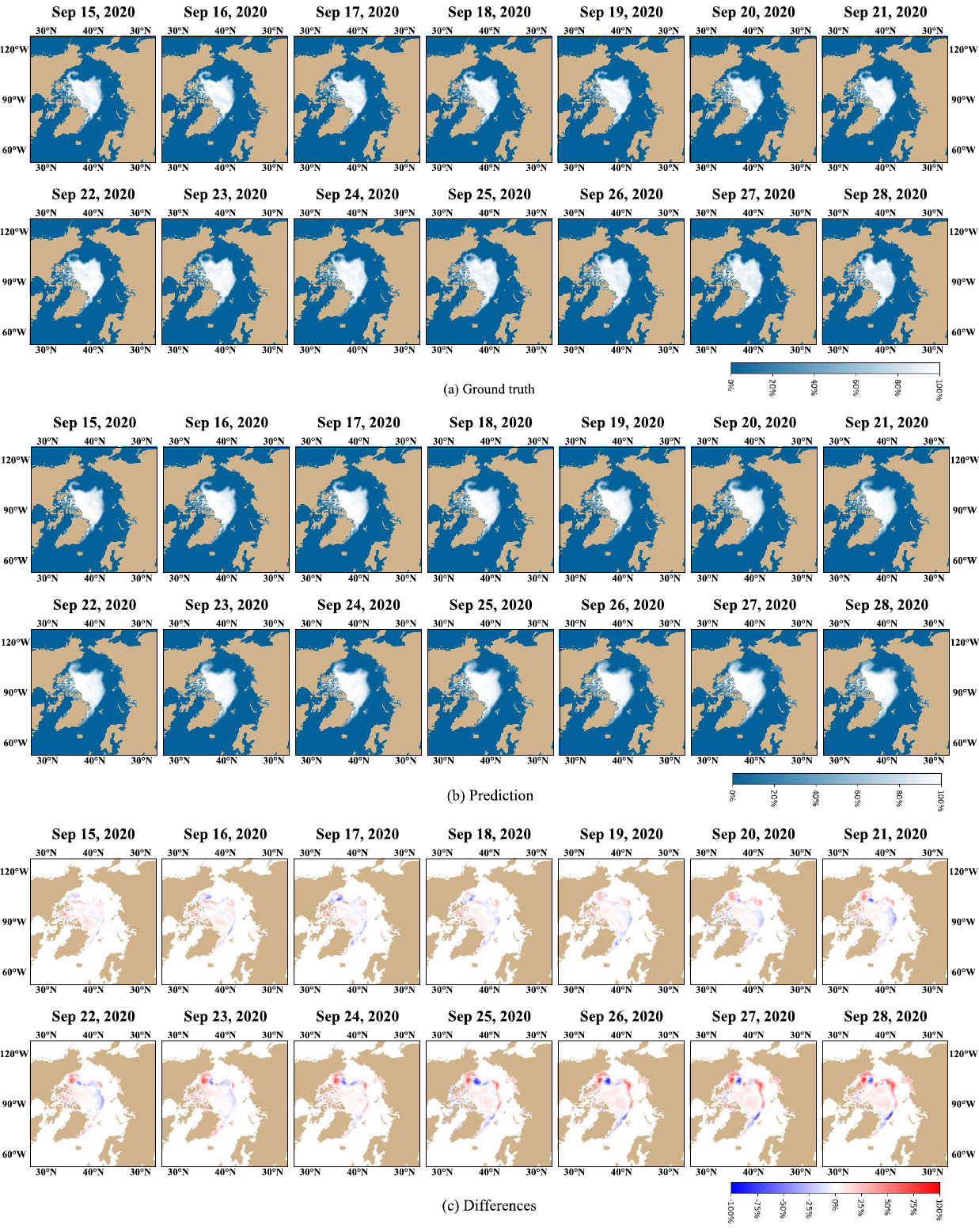}
    \caption{Details of the SIC prediction results for September 15–28, 2020. (a) Ground Truth. (b) Predictions. (c) Differences.}
    \label{fig_vis2020}
\end{figure*}

\textbf{Implementation Details.} In the proposed FCNet architecture, the Adaptive Fourier Frequency Block (AFFB) is iteratively employed eight times to efficiently extract and refine features within the frequency domain. Concurrently, the convolutional feature extraction branch comprises three encoder layers and three decoder layers, facilitating comprehensive feature processing. Furthermore, the High-Frequency Enhancement Block (HFEB) is also repeated eight times to capture intricate dynamic spatial-temporal features. The dataset encompasses a broad temporal span from 1991 to 2020. Specifically, the period from 1991 to 2010 is designated as the training set, 2011 to 2015 as the validation set, and 2016 to 2020 as the test set. The model was implemented using PyTorch 2.4.1 and trained on an NVIDIA® GeForce RTX 4090 GPU, with a total training duration of 8 hours. For optimization, we employed the AdamW optimizer, initiating with a learning rate of 0.001. During the training phase, OneCycleLR was utilized to dynamically adjust the learning rate, thereby enhancing model performance.

\begin{table}[]
    \centering
    \caption{Prediction results of the proposed FCNet (14 days $\rightarrow $ 14 days) from 2016 to 2020.}
    \scalebox{0.9}{
        \begin{tabular}{cccc}
            \hline\toprule
            ~~~ Year ~~~ & ~~ MAE $\downarrow$ ~~ & ~~ RMSE $\downarrow$ ~~ & ~~ NSE $\uparrow$ ~~ \\
            \midrule
            2016         & 2.327\%                & 7.046\%                 & 0.96077              \\
            2017         & 2.131\%                & 6.527\%                 & 0.96880              \\
            2018         & 2.093\%                & 6.658\%                 & 0.96873              \\
            2019         & 2.024\%                & 6.227\%                 & 0.97102              \\
            2020         & 2.068\%                & 6.470\%                 & 0.96727              \\
            \midrule
            Average      & 2.128\%                & 6.585\%                 & 0.96732              \\
            \bottomrule\hline
        \end{tabular}}
    \label{table_fcnetres}
\end{table}

\begin{figure}[htbp]
    \centering
    \subfigure[2016]{\includegraphics[width=1.6in]{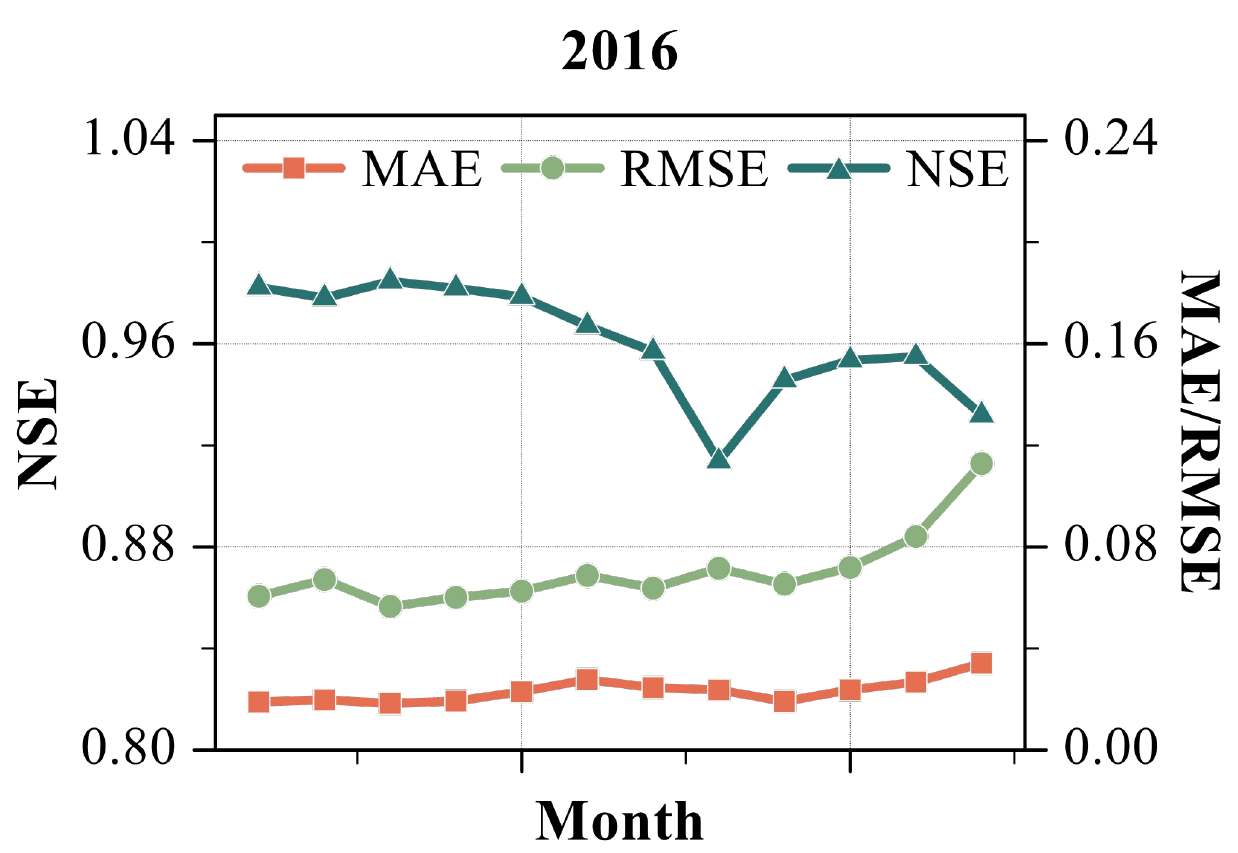}}
    \subfigure[2017]{\includegraphics[width=1.6in]{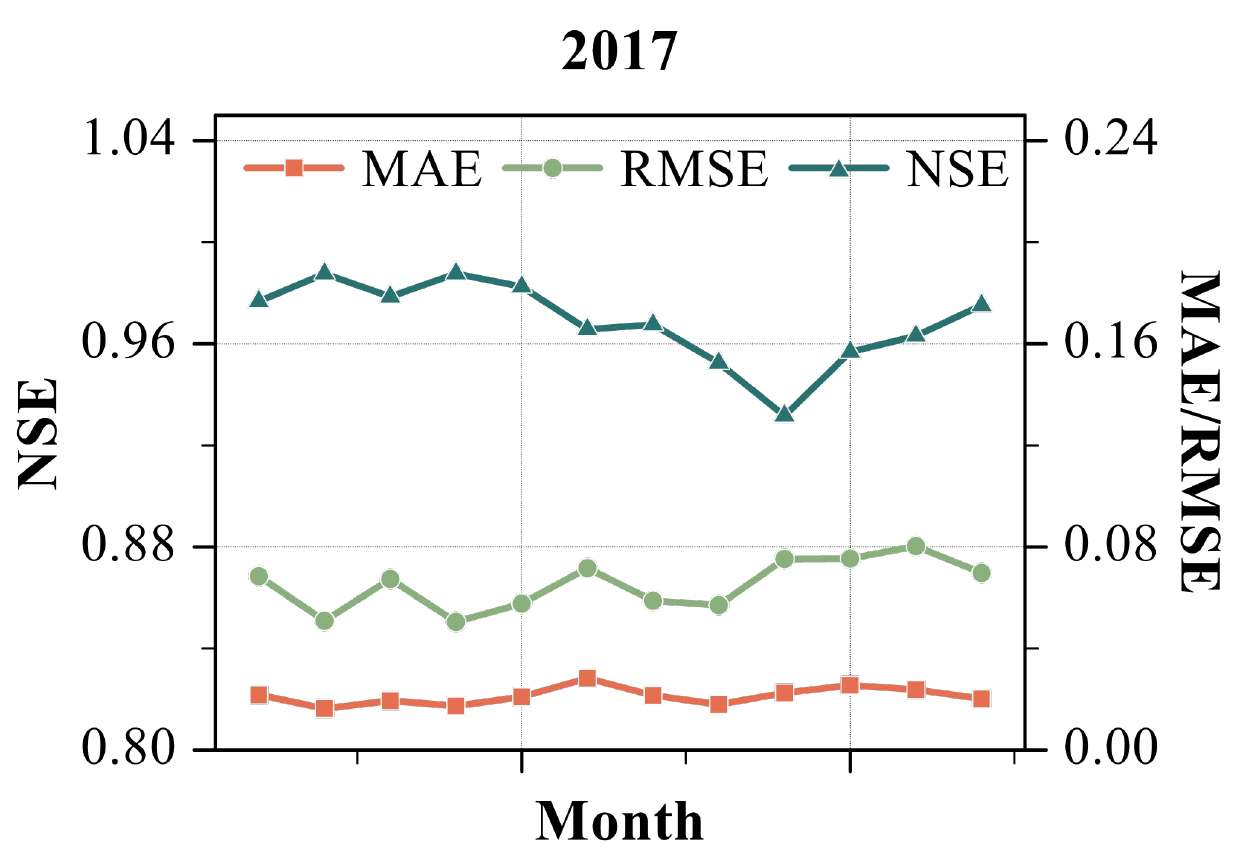}}
    \subfigure[2018]{\includegraphics[width=1.6in]{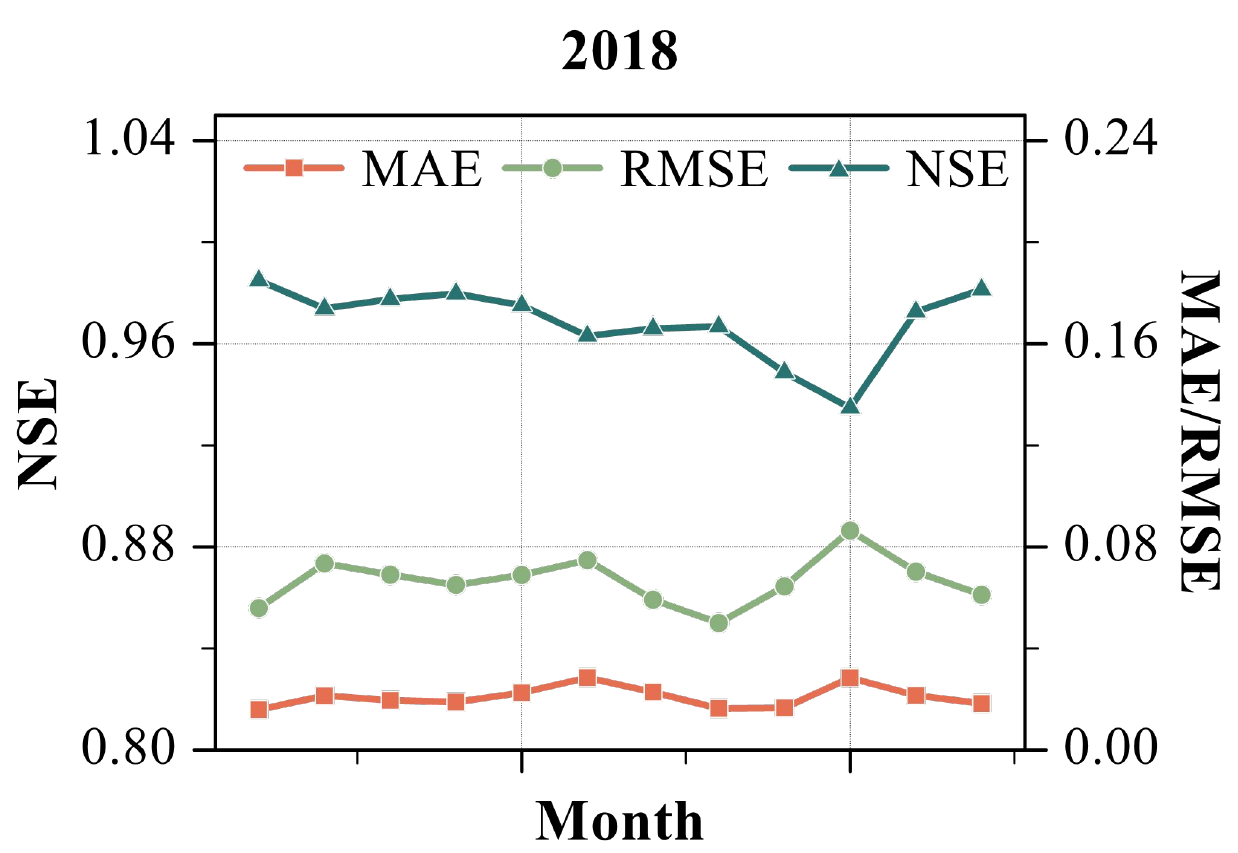}}
    \subfigure[2019]{\includegraphics[width=1.6in]{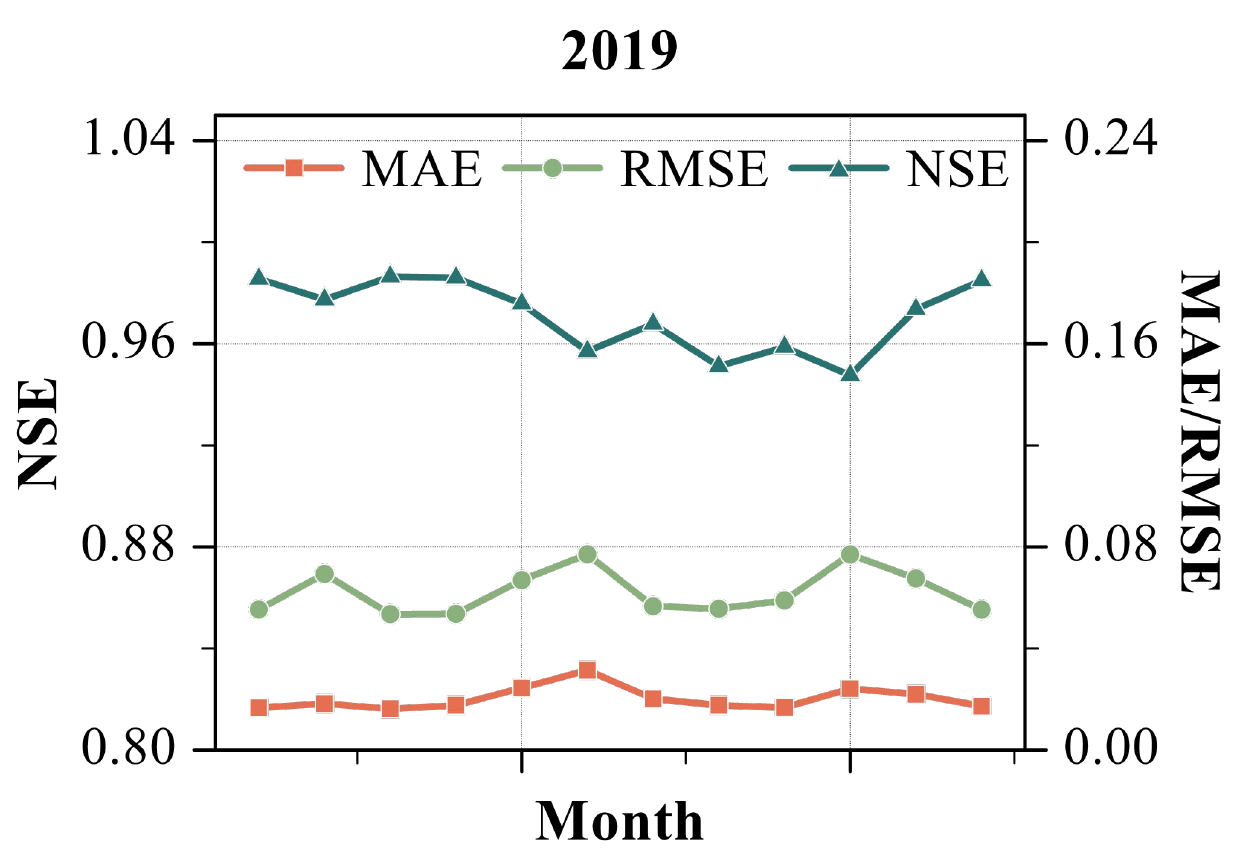}}
    \subfigure[2020]{\includegraphics[width=1.6in]{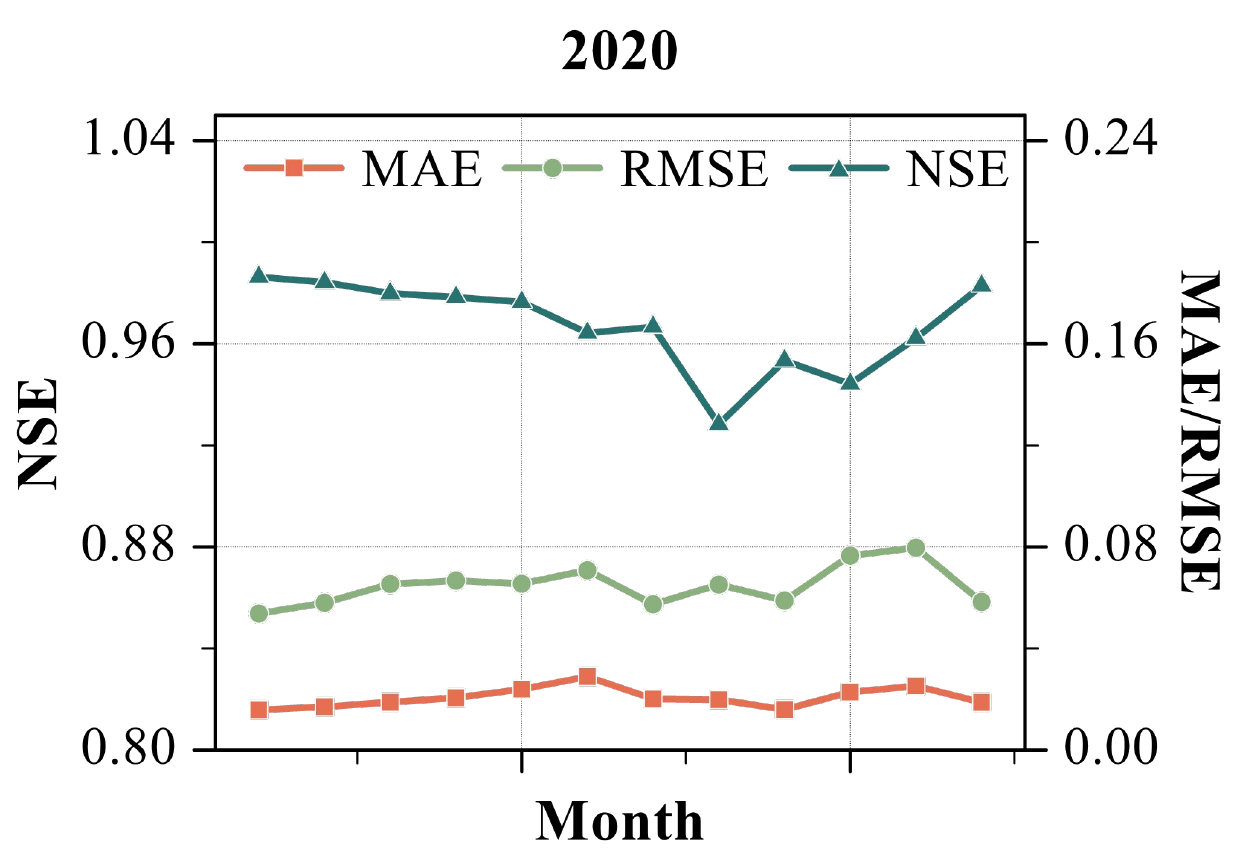}}
    \caption{Performance of our FCNet on the test set. Temporal distribution of MAE, RMSE, and NSE from 2016 to 2020.}
    \label{fig_fcnetres}
\end{figure}
\subsection{Overall Performance of the proposed FCNet}

Table \ref{table_fcnetres} shows the quantitative predictive results of the proposed FCNet on the dataset from 2016 to 2020. There is slight variations across different years, with the maximum MAE and RMSE recorded as 2.327\% (in 2016) and 7.046\% (in 2016), respectively. The lowest NSE value is 0.96077, also observed in 2016. The overall average results for MAE, RMSE, and NSE are 2.128\%, 6.585\%, and 0.96732, respectively. The experimental results demonstrate the significant predictive accuracy of FCNet, which affirms the robust performance of FCNet in SIC forecasting.

The marginally elevated forecast error observed in 2016 can be attributed to anomalous Arctic climate conditions during that period. Notably, the concurrent occurrence of unusually high sea ice concentration and elevated sea surface temperatures throughout the winter-spring season of 2016, coupled with the presence of thinner ice compared to preceding years, collectively introduced substantial uncertainties in the summer sea ice extent projections \cite{tc-12-433-2018}.

Additionally, we illustrate the temporal distribution of the monthly mean values of MAE, RMSE, and NSE in Fig. \ref{fig_fcnetres}. The MAE ranges from 1.586\% to 3.416\%, RMSE ranges from 5.004\% to 11.287\%, and NSE ranges from 0.91371 to 0.98764. Our FCNet precisely aligns with the melting process and the subsequent reverse trend from thawing to freezing. Typically, the minimum extent of Arctic sea ice occurs between September and October, indicating that the most drastic changes in sea ice take place during this period. Consequently, the NSE values are relatively low, while the MAE values are relatively high; however, they still remain at an acceptable level.

\subsection{Detailed Performance of FCNet}

In order to explore the detailed performance of FCNet, we chose a two-week period and carried out extensive experiments. As the minimum sea ice extent in the Arctic typically happens in September, this period usually experiences the lowest levels of both Arctic sea ice extent and thickness throughout the year \cite{stroeve2018changing}. We provide the visualized results of ground truth, prediction results, and the residuals in September 2016 and 2020 in Fig. \ref{fig_vis2016} and \ref{fig_vis2020}, respectively.

Fig. \ref{fig_vis2020} presents detailed information from September 15 to 28, 2020. Overall, the predicted results are consistent with the ground truth data. For the first 6 days, the majority of regions exhibit residuals within the range of (-10\%, 10\%). Over the remaining 8 days, differences gradually increase, but in most cases, residuals still fall within the range of (-20\%, 20\%), as illustrated in Fig. \ref{fig_vis2020}(c). It can be observed that our FCNet overestimates the SIC in the central regions while underestimating it in the marginal seas of the Arctic Basin. Fig. \ref{fig_vis2016} shows similar trends. For most regions, the residual range is approximately (-10\%, 10\%) during the first 6 days, as depicted in Fig. \ref{fig_vis2016}(c). Over the remaining 8 days, residuals for most regions remain within the range of (-20\%, 20\%). Through both cases, it can be found that for the most dramatic sea ice changes, the proposed FCNet achieves satisfying prediction results in September. It is worth noting that the latter 8 days of the 2016 forecast have slightly higher localized errors, again due to the unusual Arctic climatic conditions in that year.The two summer storms that entered the Arctic in 2016 resulted in unusual ice melt and dispersion patterns, complicating the September SIEE forecast \cite{tc-12-433-2018}.

\subsection{Comparison With Several DL Models}

To demonstrate the superiority of our FCNet, we conducted comparative study with two state-of-the-art methods for SIC prediction: the IceNet \cite{icenet} and the Predictive Recurrent Neural Network (PredRNN) \cite{jmse2023}. The IceNet is a deep learning U-Net model, which processes data through a series of convolutional blocks with batch normalization. The PredRNN employs spatial-temporal LSTM, and incorporate 3D convolution for state transitions.

\begin{figure}[]
    \centering
    \subfigure[MAE]{
        \label{fig_comp_mae}
        \includegraphics[width=3.2in]{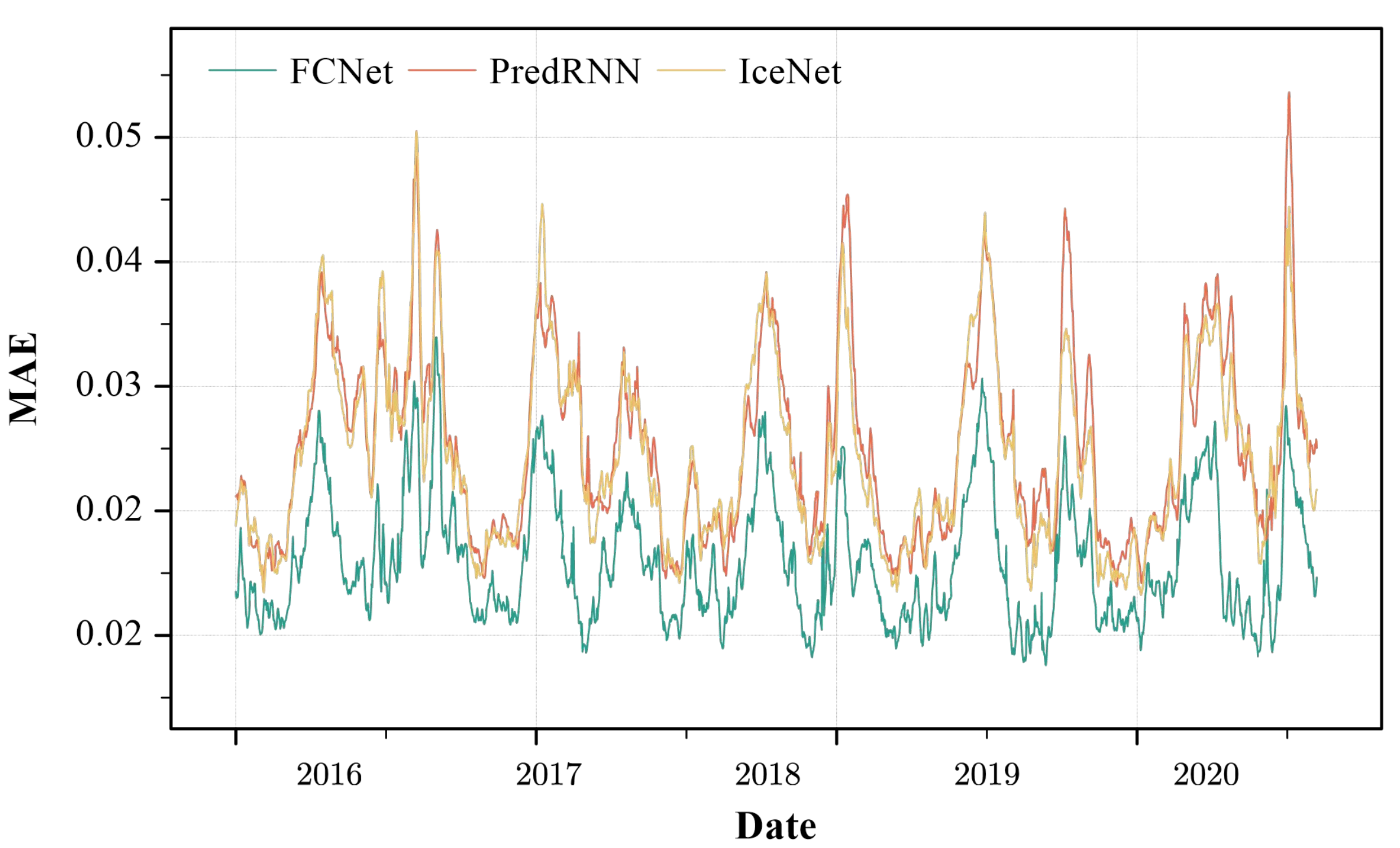}}
    \subfigure[RMSE]{
        \label{fig_comp_rmse}
        \includegraphics[width=3.2in]{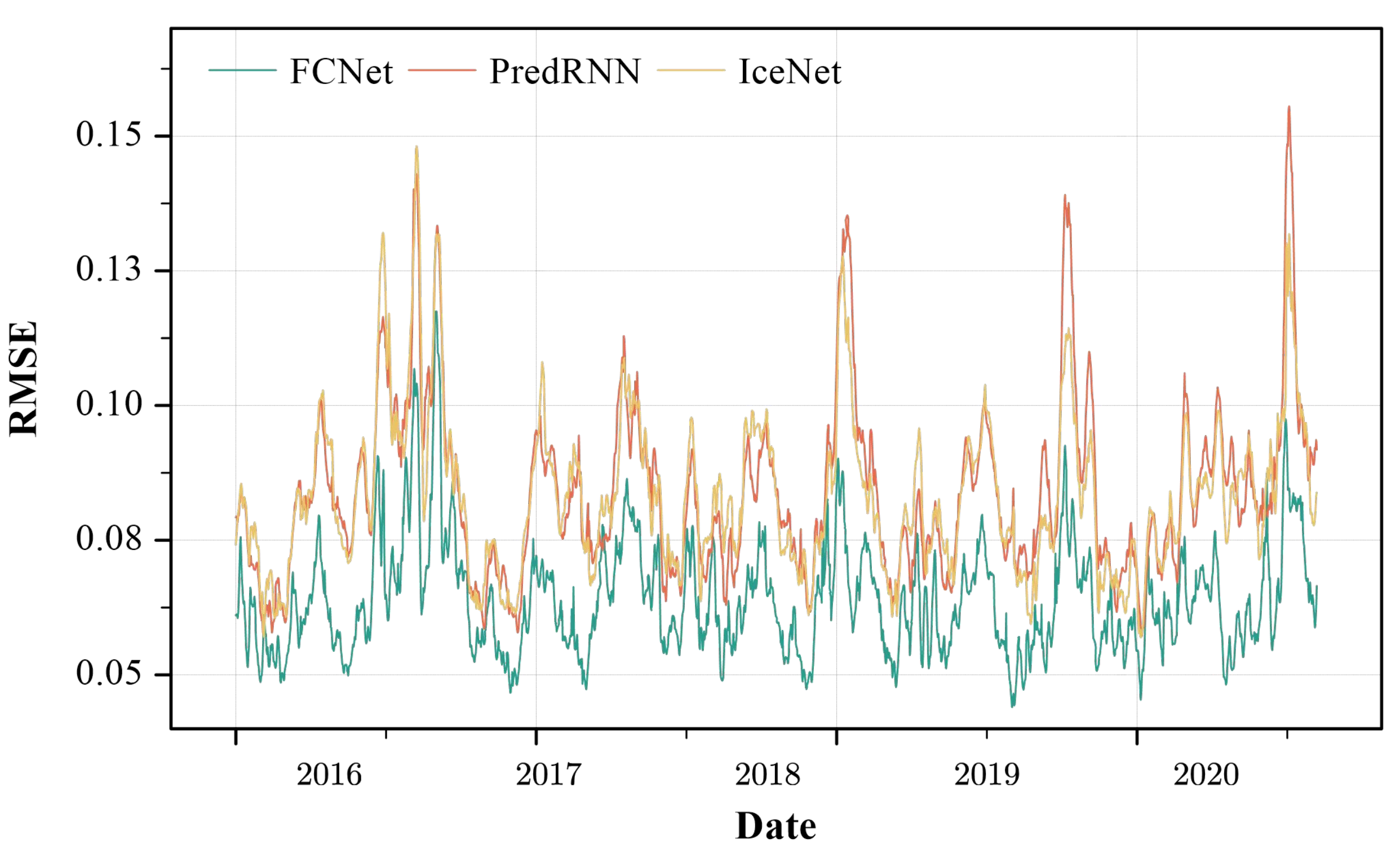}}
    \subfigure[NSE]{
        \label{fig_comp_nse}
        \includegraphics[width=3.2in]{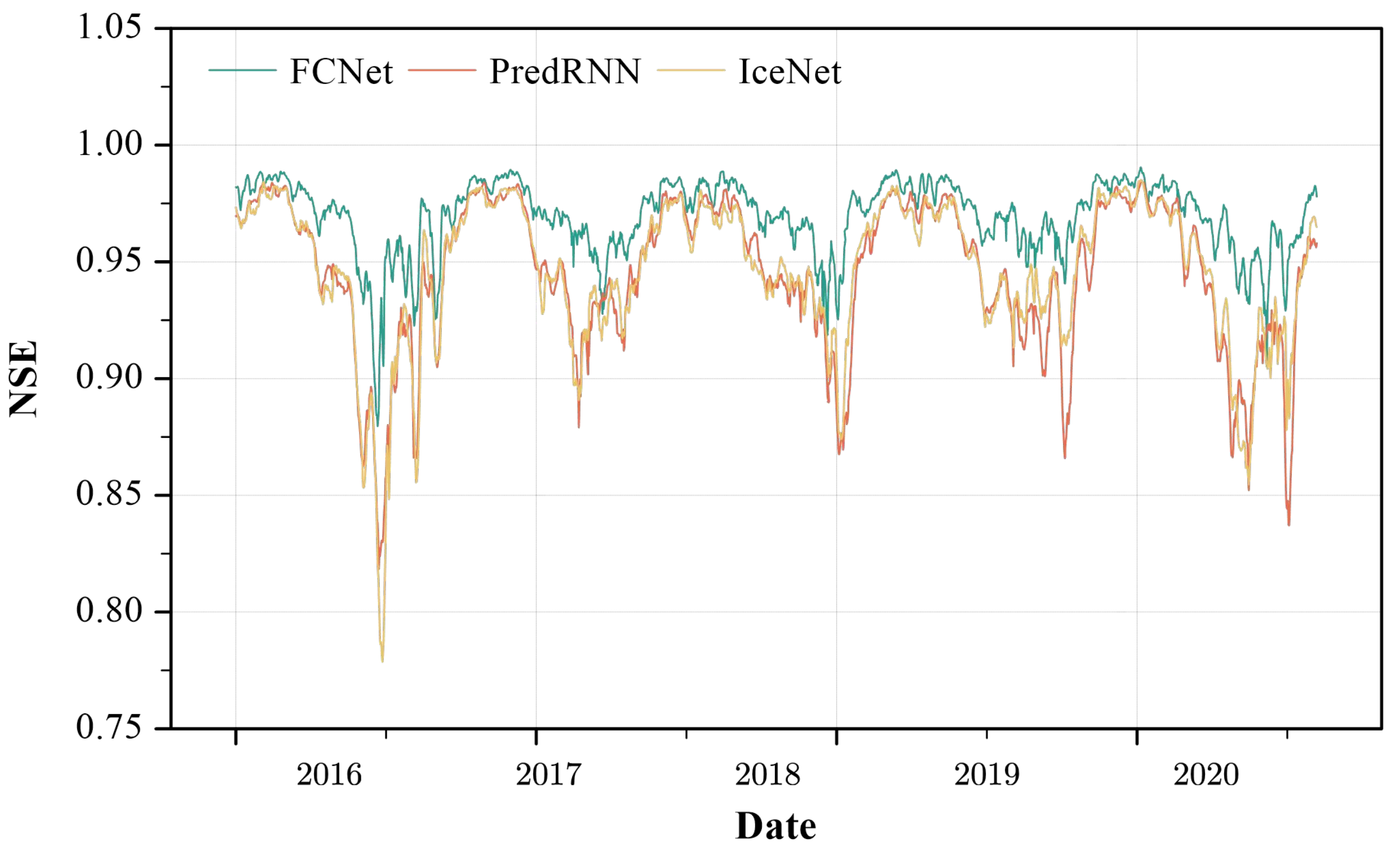}}
    \caption{Temporal distribution of MAE, RMSE, and NSE for different models from 2016 to 2020.}
    \label{fig_comp}
\end{figure}

Table \ref{table_comp} illustrates the prediction results of different methods. It is evident that the proposed FCNet outperforms other methods on all three metrics. In addition, we visualize the temporal distribution of three evaluation metrics of different methods from 2016 to 2020 in Fig. \ref{fig_comp}. It can be observed that our FCNet outperforms the other methods consistently.

\begin{table}[htbp]
    \centering
    \caption{Quantitative results of different methods on the test dataset (14 days $\rightarrow  $ 14 days) from 2016 to 2020.}
    \scalebox{0.9}{
        \begin{tabular}{cccccc}
            \multicolumn{6}{c}{MAE $\downarrow $}                                                                         \\
            \hline\toprule
            Method       & 2016             & 2017              & 2018              & 2019             & 2020             \\
            \midrule
            PredRNN      & 3.162\%          & 2.867\%           & 2.931\%           & 2.904\%          & 2.916\%          \\
            IceNet       & 3.003\%          & 2.626\%           & 2.740\%           & 2.649\%          & 2.722\%          \\
            FCNet (Ours) & \textbf{2.327\%} & \textbf{2.131\% } & \textbf{2.093\% } & \textbf{2.024\%} & \textbf{2.068\%} \\
            \bottomrule\hline

            \multicolumn{6}{c}{ }                                                                                         \\
            \multicolumn{6}{c}{RMSE $\downarrow $}                                                                        \\
            \hline\toprule
            Method       & 2016             & 2017              & 2018              & 2019             & 2020             \\
            \midrule
            PredRNN      & 9.081\%          & 8.103\%           & 8.480\%           & 8.277\%          & 8.471\%          \\
            IceNet       & 9.031\%          & 7.868\%           & 8.512\%           & 7.952\%          & 8.377\%          \\
            FCNet (Ours) & \textbf{7.046\%} & \textbf{6.527\%}  & \textbf{6.658\%}  & \textbf{6.227\%} & \textbf{6.470\%} \\
            \bottomrule\hline

            \multicolumn{6}{c}{ }                                                                                         \\
            \multicolumn{6}{c}{NSE$ \uparrow $}                                                                           \\
            \hline\toprule
            Method       & 2016             & 2017              & 2018              & 2019             & 2020             \\
            \midrule
            PredRNN      & 0.93088          & 0.95348           & 0.94735           & 0.94907          & 0.94386          \\
            IceNet       & 0.93030          & 0.95584           & 0.94623           & 0.95320          & 0.94405          \\
            FCNet (Ours) & \textbf{0.96077} & \textbf{0.96880}  & \textbf{0.96873}  & \textbf{0.97102} & \textbf{0.96727} \\
            \bottomrule\hline
        \end{tabular}}
    \label{table_comp}
\end{table}

\begin{figure*}[htbp]
    \centering
    \subfigure[Differences between PredRNN and our FCNet ($\textrm{MAE}_\textrm{PredRNN}-\textrm{MAE}_\textrm{FCNet}$)]{
        \label{fig_maediff_predrnn}
        \includegraphics[width=\textwidth]{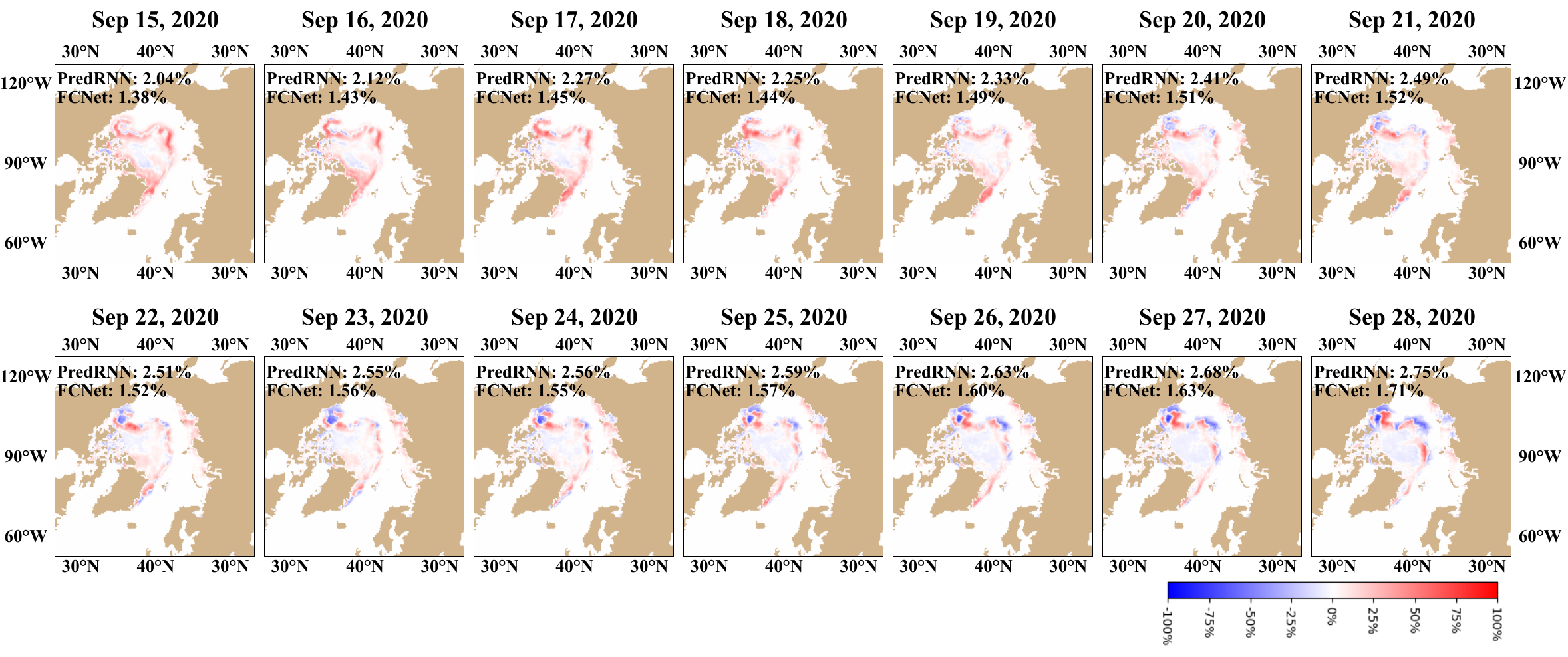}}
    \subfigure[Differences between IceNet and our FCNet ($\textrm{MAE}_\textrm{IceNet}-\textrm{MAE}_\textrm{FCNet}$)]{
        \label{fig_maediff_icenet}
        \includegraphics[width=\textwidth]{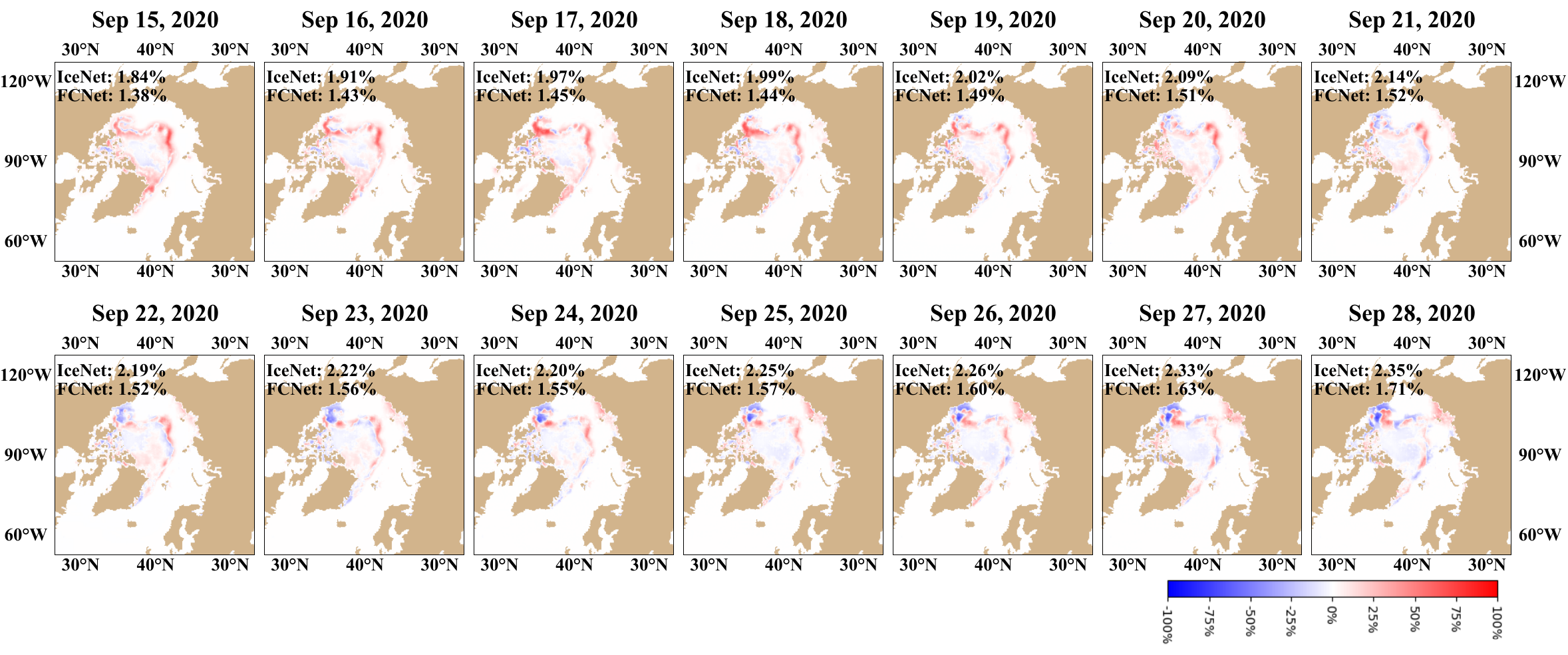}}
    \caption{Spatial comparison of MAE among the FCNet, IceNet, and PredRNN models.}
    \label{fig_maediff}
\end{figure*}

\begin{figure}[htbp]
    \centering
    \subfigure[Sep 28, 2019.]{
        \label{fig_deatils2019}
        \includegraphics[width=3.2in]{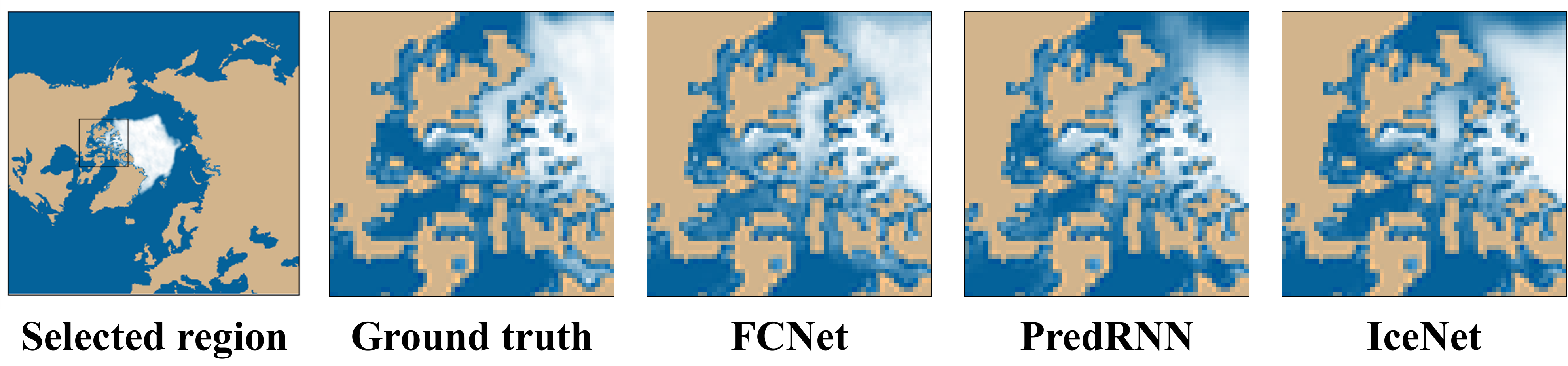}}
    \subfigure[Sep 15, 2020.]{
        \label{fig_deatils2020}
        \includegraphics[width=3.2in]{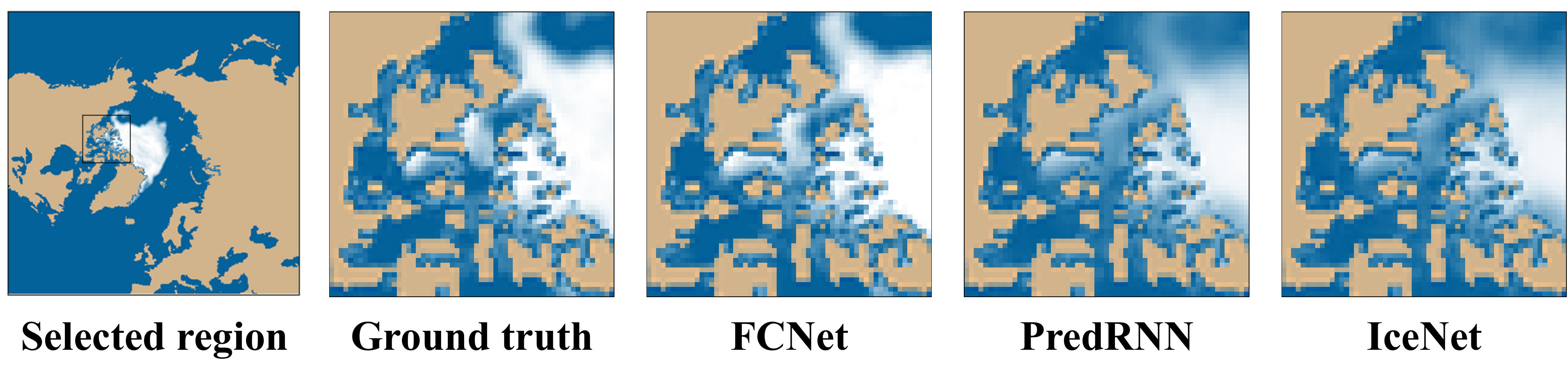}}
    \caption{Comparison of ice sheet cracks and openings among the FCNet, IceNet, and PredRNN models.}
    \label{fig_details}
\end{figure}

We visualized the spatial distribution of MAE differences among the three models in Fig. \ref{fig_maediff}. Specifically, Fig. \ref{fig_maediff_predrnn} presents the MAE differences between PredRNN and our FCNet, while Fig. \ref{fig_maediff_icenet} presents the MAE differences between IceNet and our FCNet. The red areas indicate MAE values of PredRNN/ICENet is greater than our FCNet. The blue areas denote the MAE values of PredRNN/IceNet is smaller than our FCNet. It can be observed that, for the majority of regions, the proposed FCNet outperforms PredRNN and IceNet. Furthermore, the proposed FCNet performs better in edge regions. We have selected several areas in the Arctic that exhibit relatively rich details and textures. As shown in Fig. \ref{fig_details}, our FCNet can better predict the cracks and openings in the ice sheet. From this, we can conclude that our FCNet has a strong ability to preserve high-frequency details, as the frequency characteristic extraction branch retains edges and textures.

\begin{figure*}[]
    \centering
    \subfigure[BACC of daily SIC prediction.]{
        \label{BACC_SEAS5}
        \includegraphics[width=3.2in]{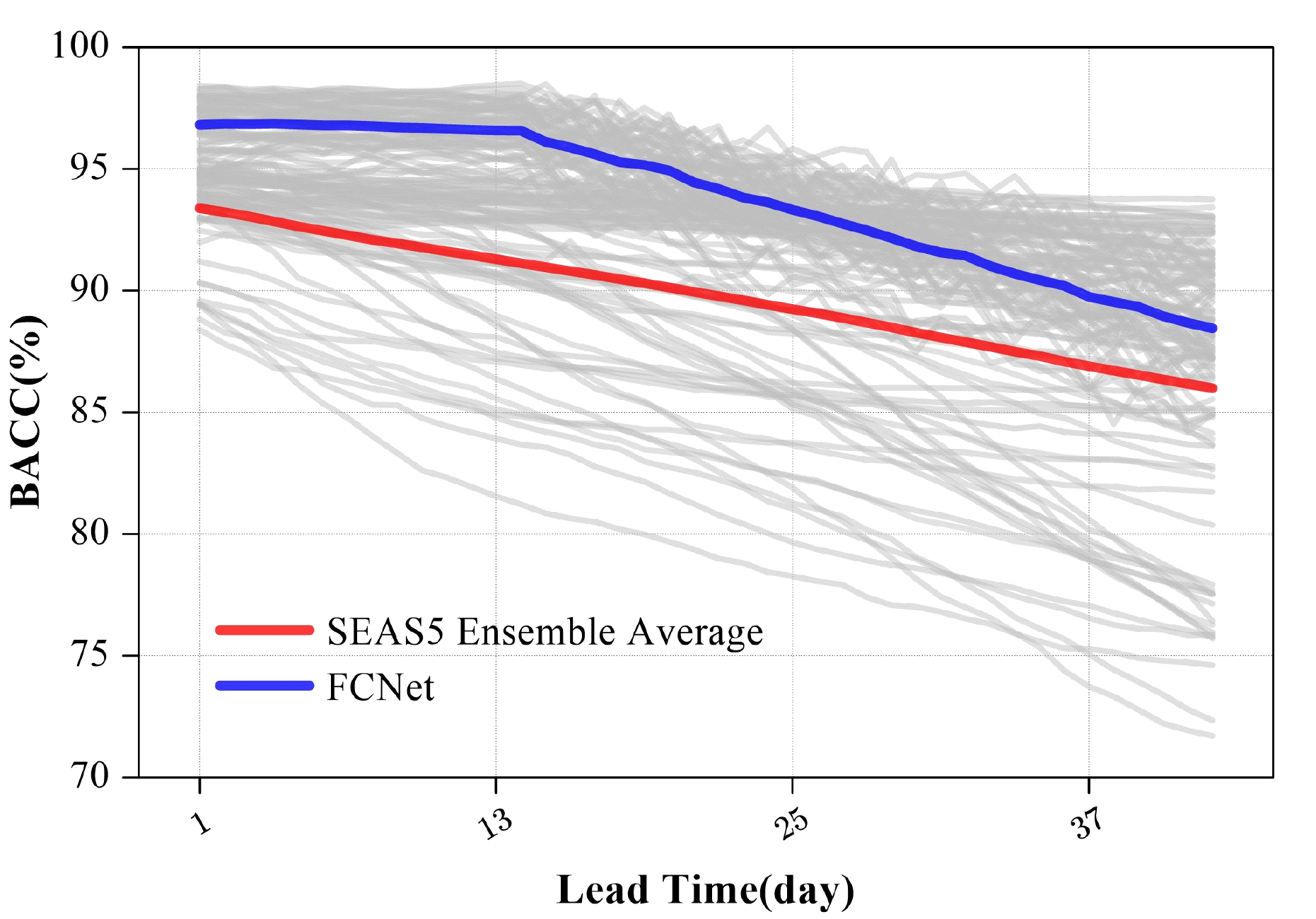}}
    \subfigure[Monthly average BACC of SIC prediction.]{
        \label{BACC_SEAS5_Monthly}
        \includegraphics[width=3.2in]{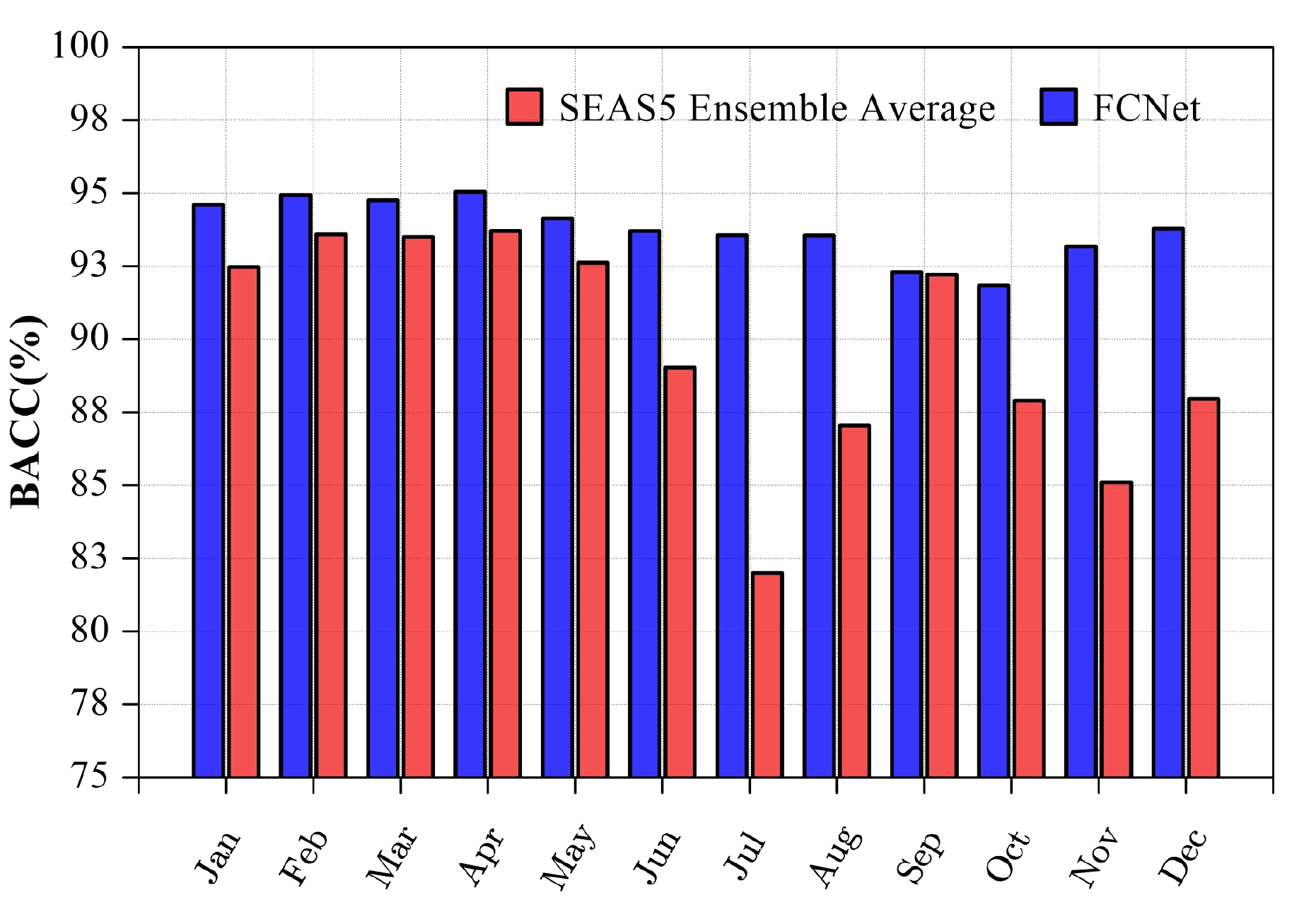}}
    \caption{Comparison of prediction performance of FCNet and SEAS5.}
    \label{fig_seas5}
\end{figure*}

\begin{figure*}
    \centering
    \label{fig_seas5_fcnet}
    \includegraphics[width=6.4in]{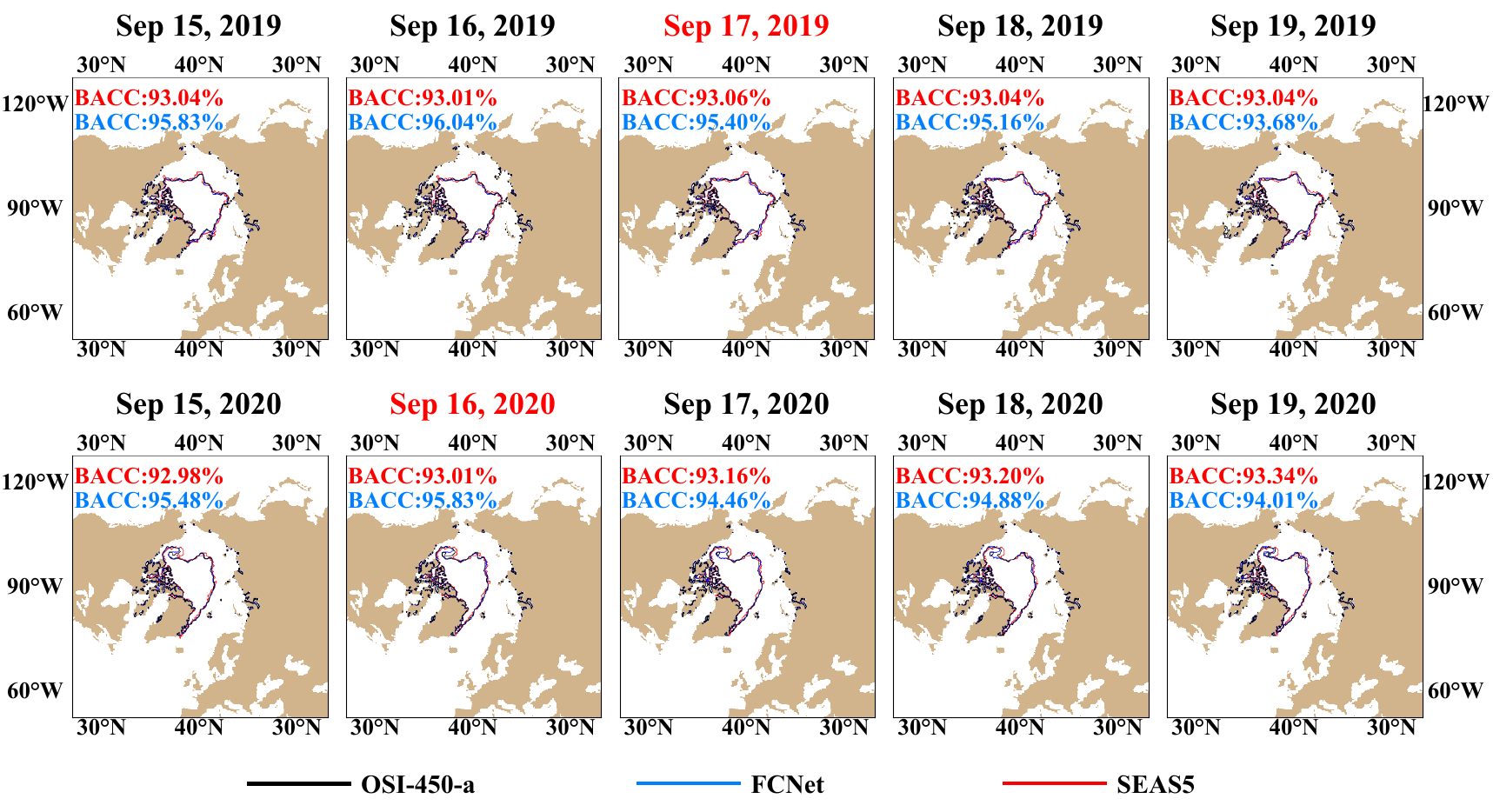}
    \caption{BACC and SIE of FCNet and SEAS5 (2019 and 2020).}
\end{figure*}

\subsection{Comparison With Numerical Model}

It is well known that Arctic sea ice prediction relies heavily on coupled ice-sea numerical models. In order to better assess the forecasting performance of FCNet, we compare its results with the daily forecast SIC of ECMWF's seasonal forecasting system SEAS5 \cite{copernicus_climate_change_service_2018}, which consists of 51 members generated by adding different small perturbations to the initial state and internal parameters of the numerical model.The SEAS5 ensemble of forecasts can more efficiently compensate for the uncertainty of a single model. Considering that SEAS5 forecasts are released only on the first day of each month, we extracted a total of 60 forecast sets (days 1-42) for the period from 2016 to 2020. Similarly, we configured FCNet to launch forecasts on the first day of each month.

The ensemble forecasting capability of SEAS5 can effectively mitigate the uncertainties associated with a standalone model. As illustrated in Fig. \ref{BACC_SEAS5}, the BACC for the ensemble-averaged SIC surpasses 0.7 with a forecast lead time of 42 days. Notably, while the trend in BACC observed for FCNet mirrors that of SEAS5, the average BACC of FCNet consistently outperforms the ensemble average of SEAS5. This superiority stems from the numerical model's limitation to a single time step per input/output operation, which accelerates and exacerbates the accumulation of errors. As shown in Fig. \ref{BACC_SEAS5_Monthly}, FCNet outperforms SEAS5 in predicting the monthly average SIC from June to December. It is worth mentioning that the monthly average BACC of FCNet is almost stable around 0.9, which suggests that FCNet is insensitive to seasons and has strong robustness.

To further confirm the stability of FCNet, we selected OSI SAF SIC data for September 2019 and September 2020 for comparative analyses with the predictions. It is worth noting that the smallest sea ice extent (SIE) recorded by satellite observations in 2019 occurs on 17 September 2019, while the second smallest SIE recorded by Arctic satellite observations occurs on 16 September 2020. As shown in Fig. \ref{fig_seas5_fcnet}, on 17 September 2019, the difference in BACC between the SIE predictions of FCNet and SEAS5 is 2.34\%, with both models exceeding 93\% accuracy. In contrast, on 16 September 2020, the BACC difference between the SIE predictions of FCNet and SEAS5 is 2.82\%, and it is noteworthy that FCNet shows superior performance throughout.

\begin{table}[htbp]
    \centering
    \caption{Ablation study of our FCNet.}
    \scalebox{0.9}{
        \begin{tabular}{cccc}
            \hline\toprule
            ~~~~ Model ~~~~                      & ~ MAE $\downarrow$ ~ & ~ RMSE $\downarrow$ ~ & NSE $\uparrow$   \\
            \midrule
            FCNet w/o AFFB                       & 2.331\%              & 6.955\%               & 0.95838          \\
            FCNet w/o HFEB                       & 2.638\%              & 7.489\%               & 0.94324          \\
            FCNet w/o $\mathcal{L}_{freq}$       & 2.292\%              & 6.729\%               & 0.96027          \\
            FCNet w/o AFFB\&$\mathcal{L}_{freq}$ & 2.394\%              & 7.105\%               & 0.95213          \\
            FCNet w/o HFEB\&$\mathcal{L}_{freq}$ & 2.763\%              & 7.768\%               & 0.94008          \\
            Vision Transformer                   & 2.766\%              & 7.834\%               & 0.93905          \\
            FCNet                                & \textbf{2.128\%}     & \textbf{6.585\%}      & \textbf{0.96732} \\
            \bottomrule\hline
        \end{tabular}}
    \label{table_ablation}
\end{table}

\begin{figure*}[htbp]
    \centering
    \subfigure[Spatial distributions of MAE differences(FCNet w/o AFFB - FCNet).]{
        \label{fig_FCNet_wo_AFFB}
        \includegraphics[width=\textwidth]{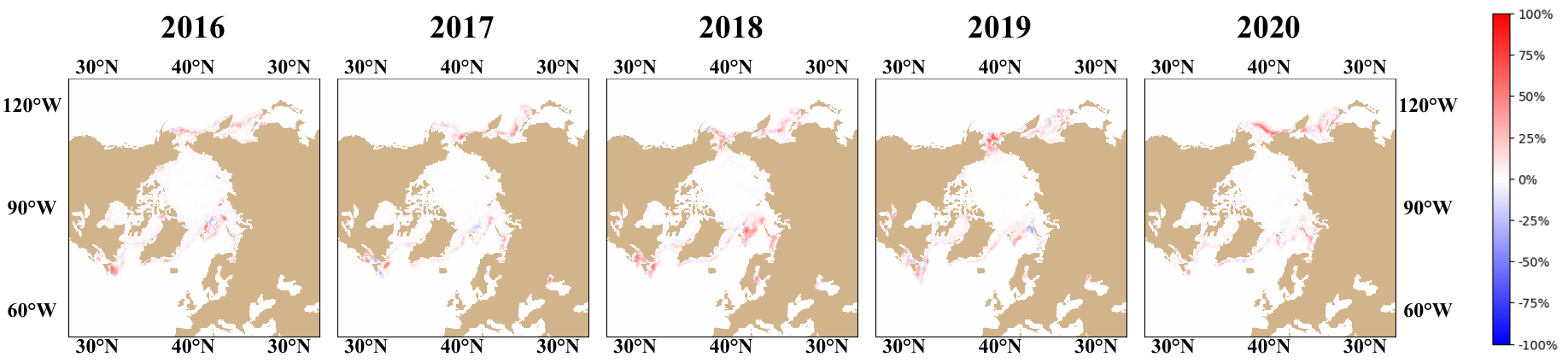}}
    \subfigure[Spatial distributions of MAE differences(FCNet w/o HFEB - FCNet).]{
        \label{fig_FCNet_wo_HFEB}
        \includegraphics[width=\textwidth]{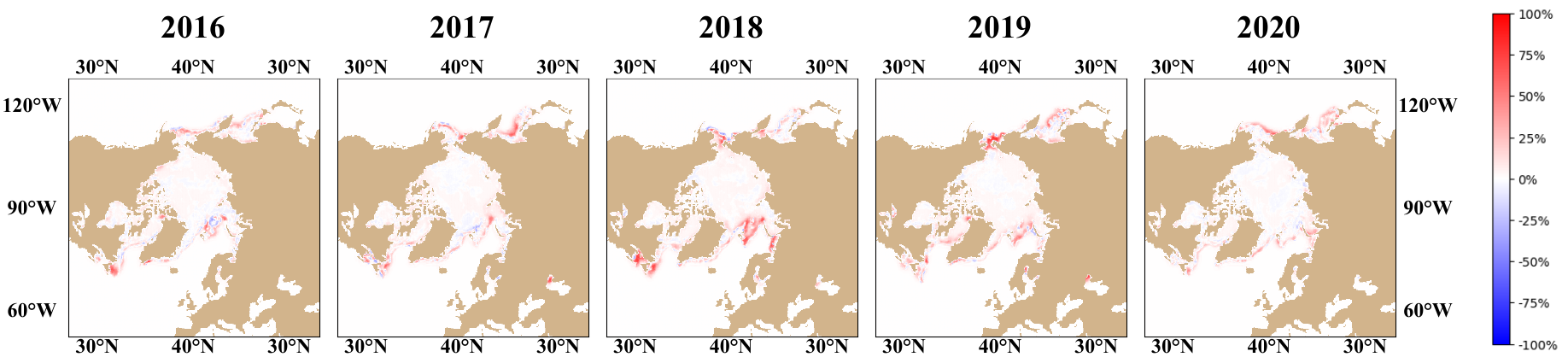}}
    \caption{Spatial distributions of MAE differences among FCNet w/o AFFB, FCNet w/o HFEB and FCNet.}
    \label{fig_ablation}
\end{figure*}

\begin{table}[htbp]
    \centering
    \caption{Performances of FCNet with different prediction period.}
    \scalebox{0.9}{
        \begin{tabular}{ccccc}
            \hline\toprule
            Recursive step & Lead time & MAE $\downarrow $ & RMSE $\downarrow $ & NSE $\uparrow $ \\
            \midrule
            0              & 14 days   & 2.128\%           & 6.585\%            & 0.96732         \\
            1              & 28 days   & 2.786\%           & 8.551\%            & 0.95676         \\
            2              & 42 days   & 3.042\%           & 9.037\%            & 0.94251         \\
            3              & 56 days   & 4.207\%           & 12.723\%           & 0.92552         \\
            \bottomrule\hline
        \end{tabular}}
    \label{table_recur}
\end{table}

\begin{figure*}[htbp]
    \centering
    \subfigure[MAE]{
        \includegraphics[width=3.2in]{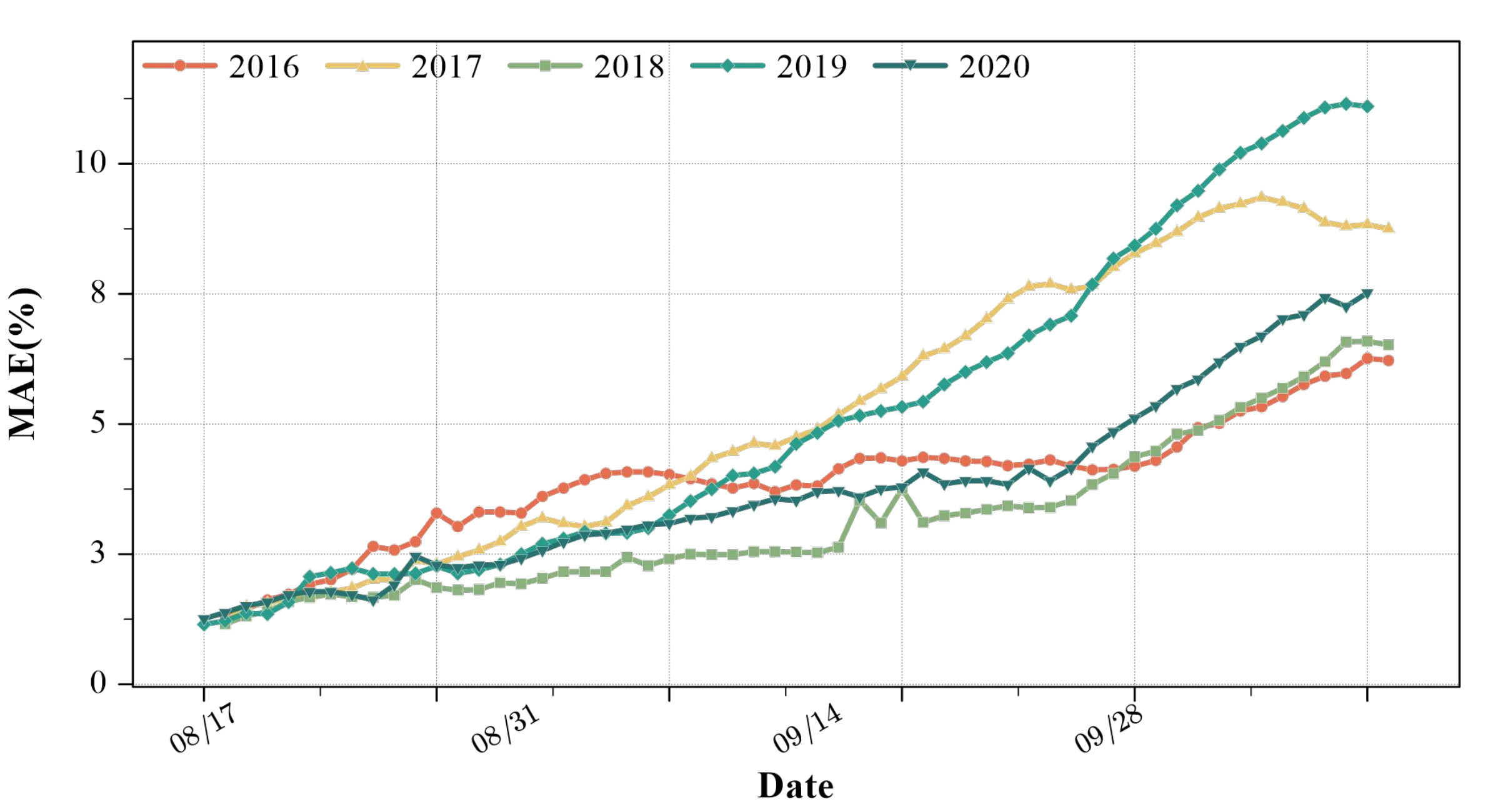}
        \label{fig_recur_mae}}
    \subfigure[RMSE]{
        \includegraphics[width=3.2in]{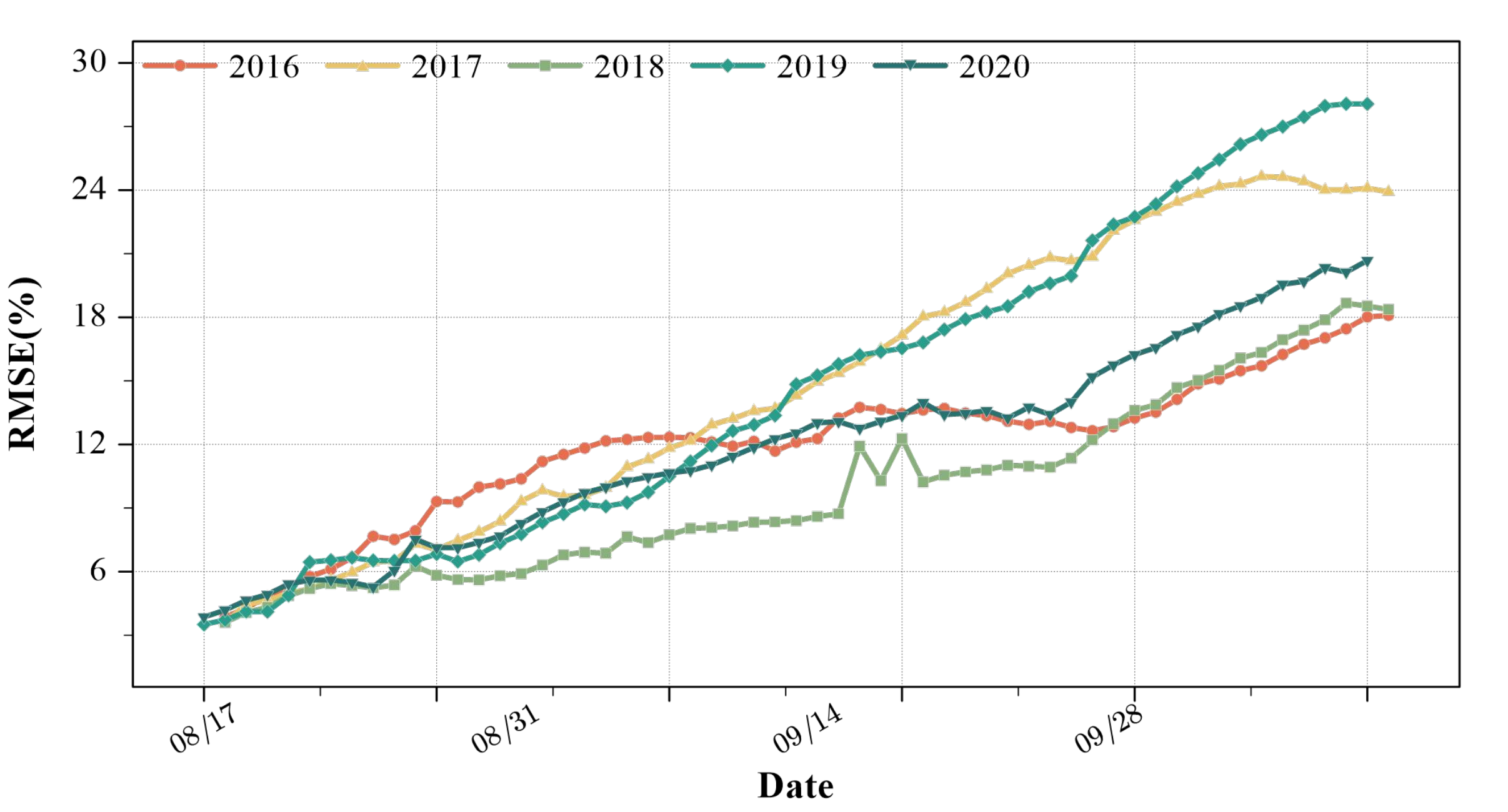}
        \label{fig_recur_rmse}}
    \subfigure[NSE]{
        \includegraphics[width=3.2in]{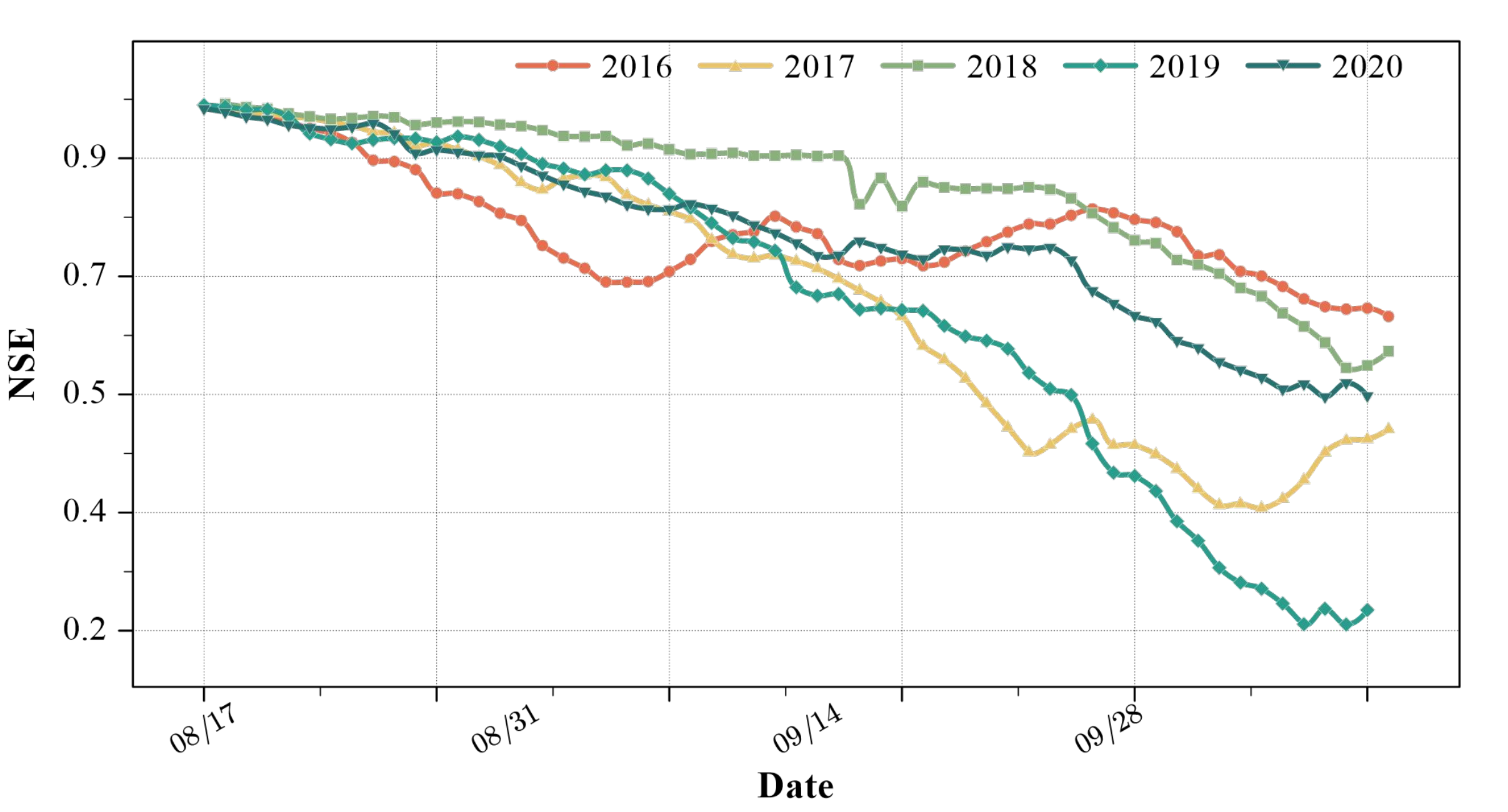}
        \label{fig_recur_nse}}
    \subfigure[BACC]{
        \includegraphics[width=3.2in]{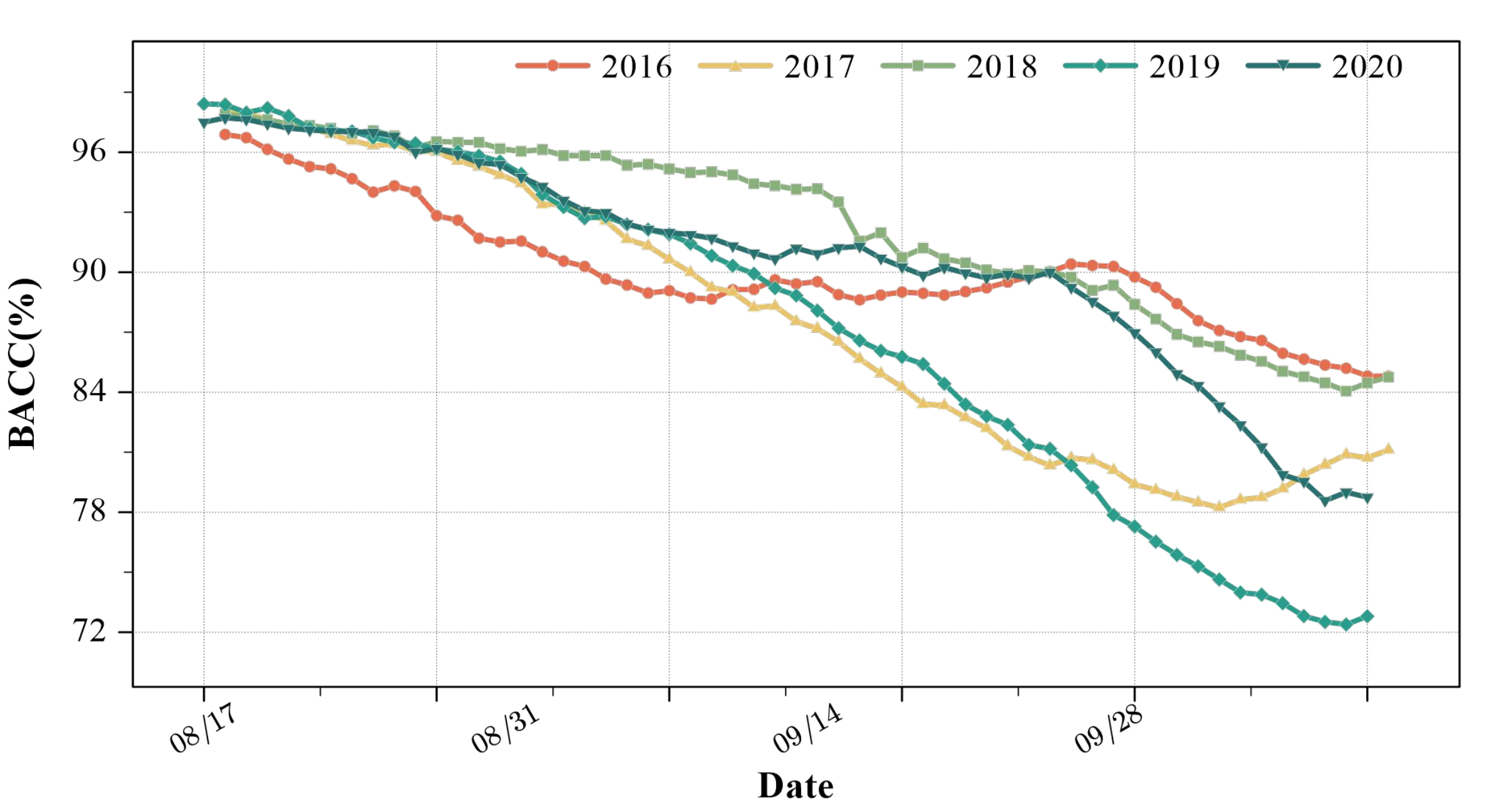}
        \label{fig_recur_bacc}}
    \caption{Experimental results of our FCNet for recursive predictions from 2016 to 2020.}
    \label{fig_recur_metric}
\end{figure*}

\begin{figure*}[htbp]
    \centering
    \includegraphics [width=6.4in]{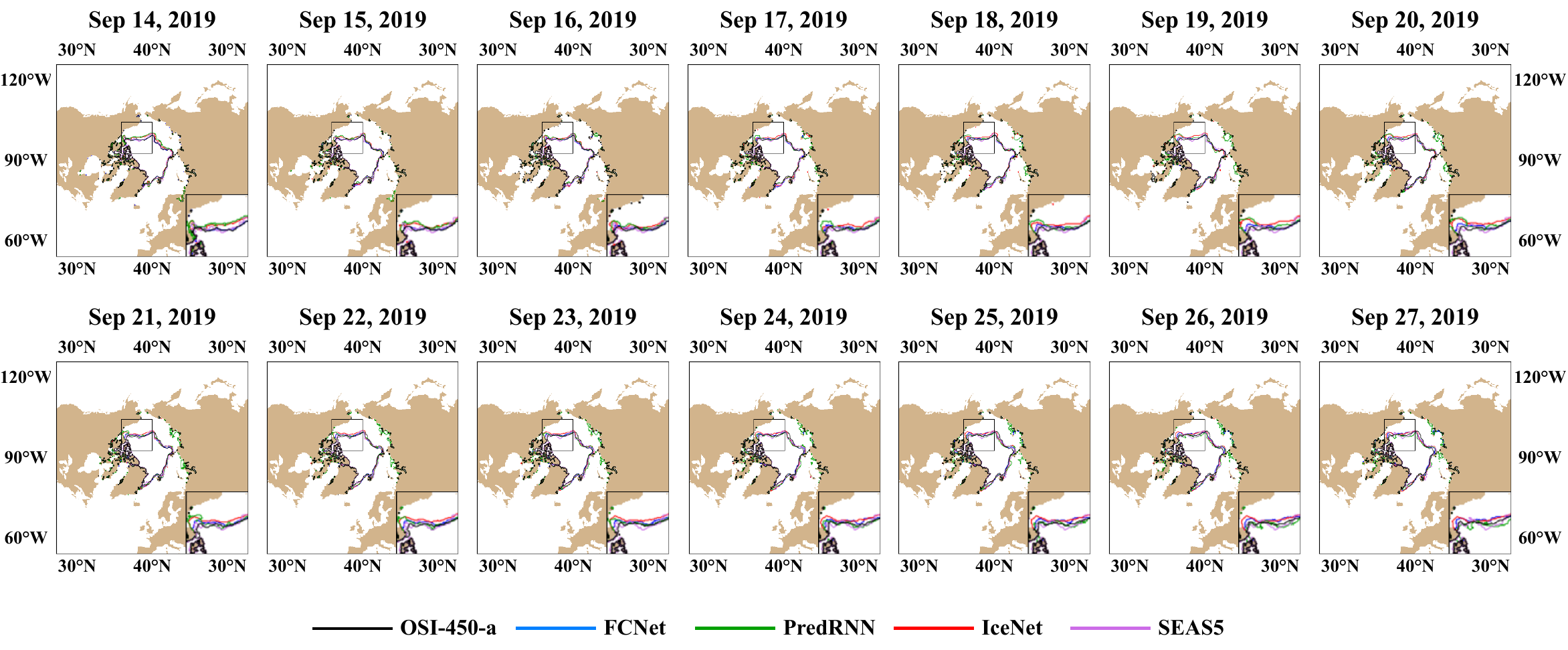}
    \caption{Sea ice extent of different methods for the period of September 14-27, 2019.}
    \label{fig_sie}
\end{figure*}

\begin{figure*}[htbp]
    \centering
    \subfigure[2007.]{
        \label{fig_case2007}
        \includegraphics[width=0.9\textwidth]{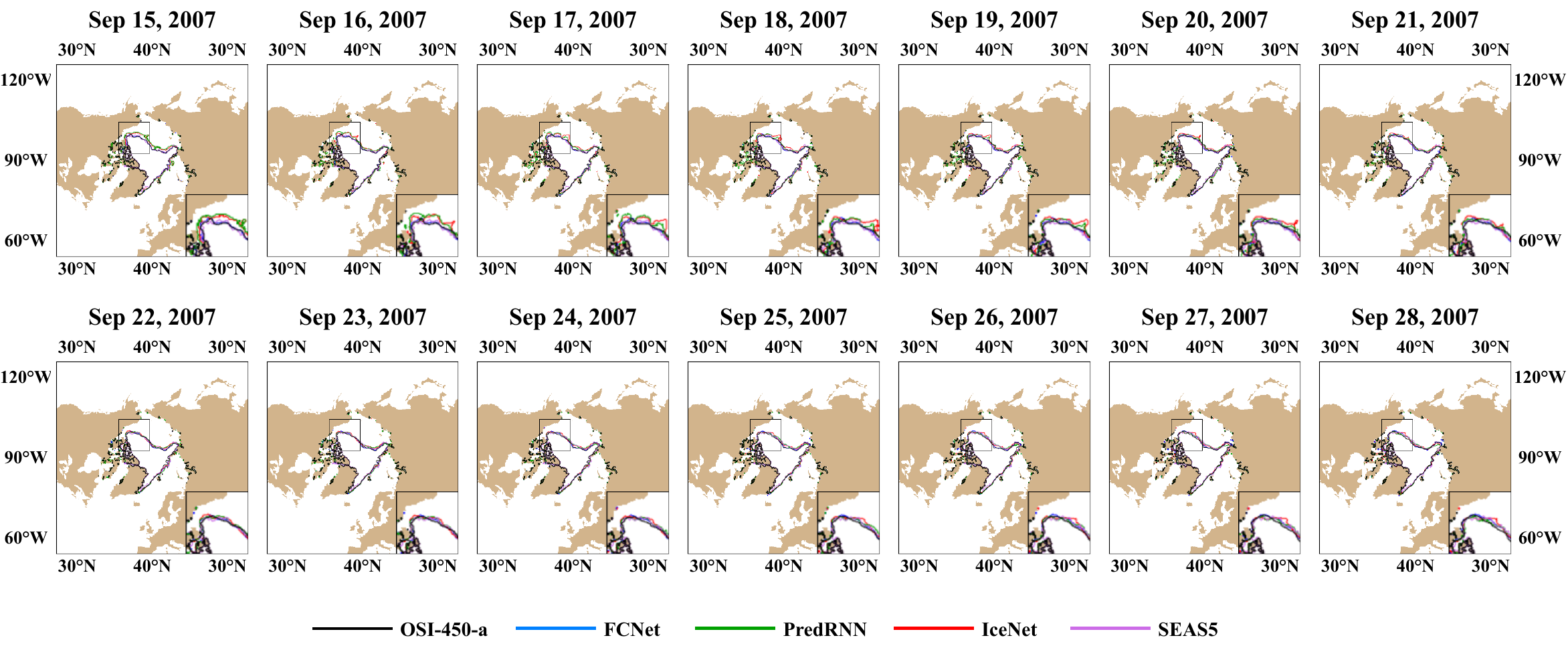}}
    \subfigure[2012.]{
        \label{fig_case2012}
        \includegraphics[width=0.9\textwidth]{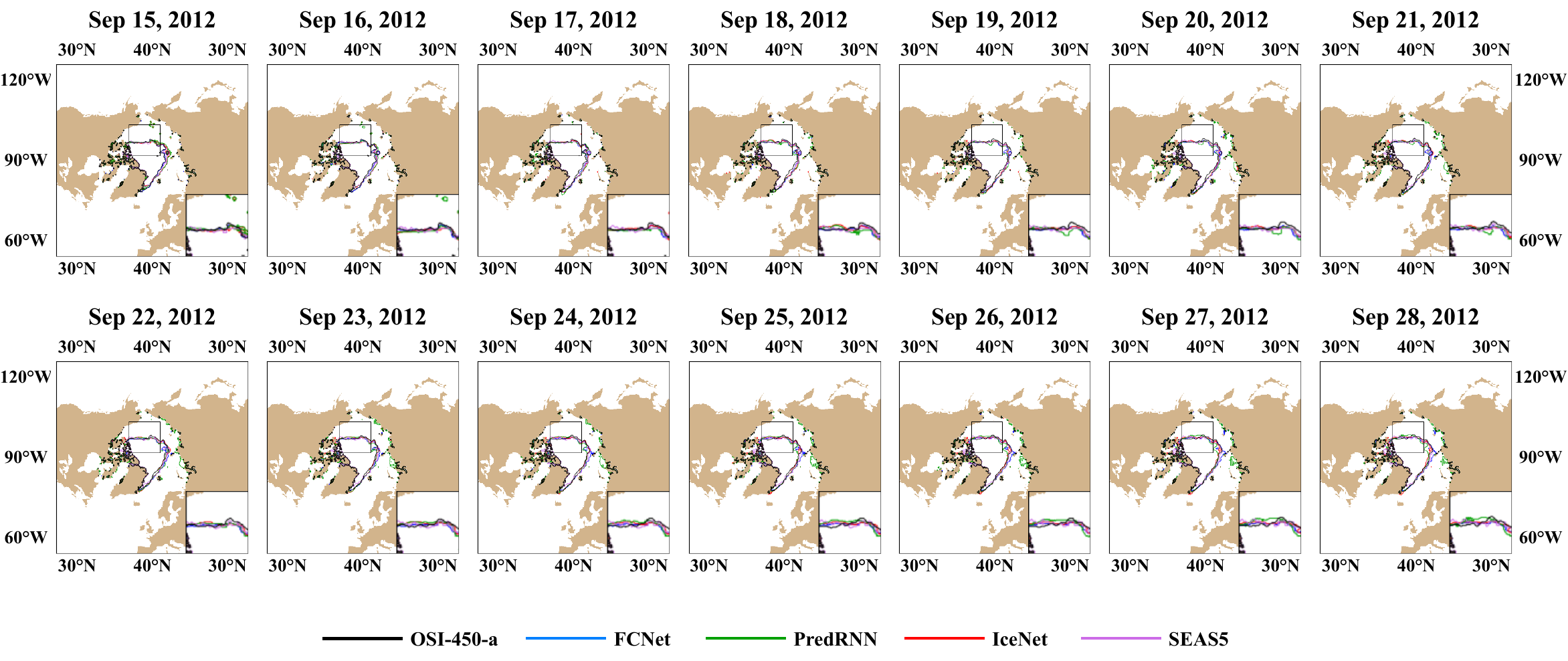}}
    \subfigure[2020.]{
        \label{fig_case2020}
        \includegraphics[width=0.9\textwidth]{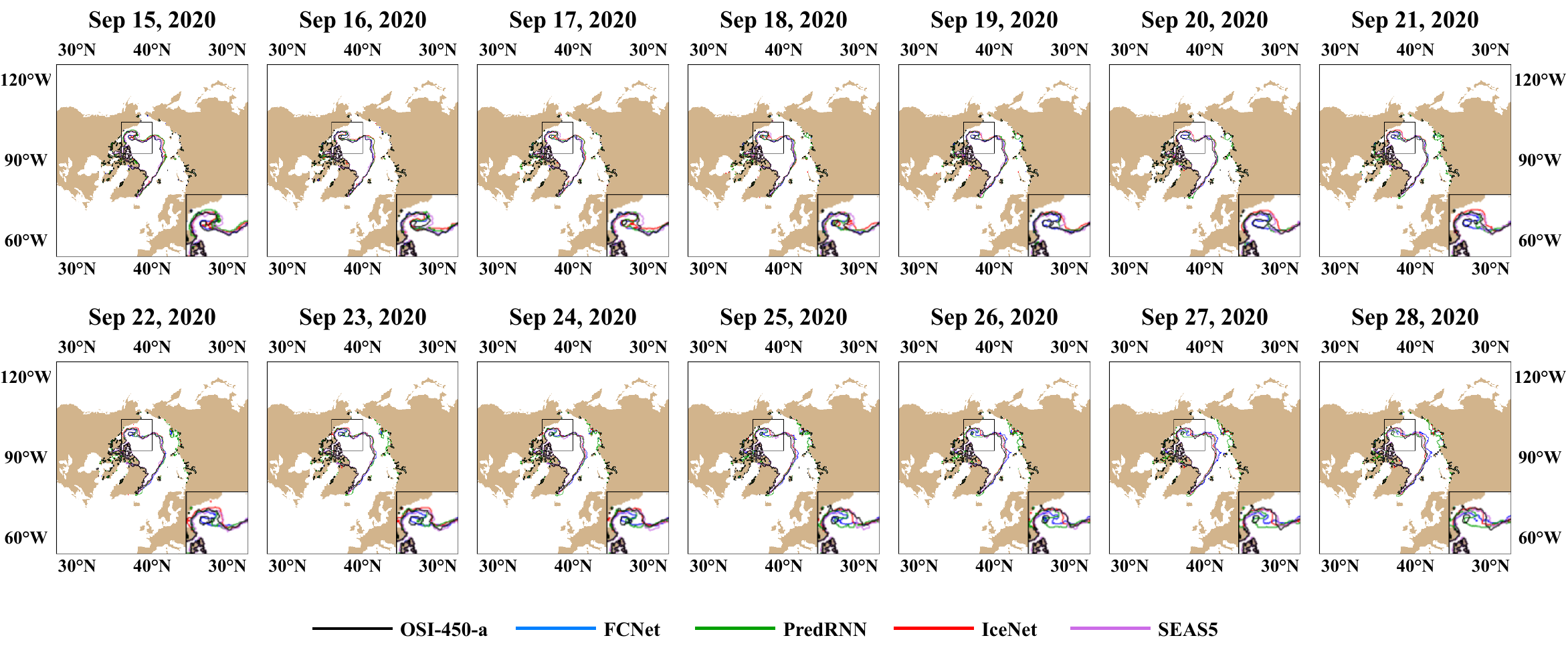}}
    \caption{Sea ice extent predicted by different methods for September 15-28 in 2007, 2012, and 2020, respectively.}
    \label{fig_extremecase}
\end{figure*}

\subsection{Ablation Study}

To verify the effectiveness of critical components of the proposed FCNet, seven models were compared: FCNet w/o AFFB, FCNet w/o HFEB, FCNet w/o $\mathcal{L}_\textrm{freq}$, FCNet w/o AFFB\&$\mathcal{L}_{freq}$, FCNet w/o HFEB\&$\mathcal{L}_{freq}$, a Vision Transformer baseline and the full FCNet (`w/o' means `without'). The experimental data and settings remain unchanged. As shown in Table \ref{table_ablation},  the removal of the AFFB resulted in a 0.403\% increase in MAE, indicating that the AFFB significantly enhances the model's ability to extract and utilize frequency features, thereby improving prediction accuracy. Furthermore, the ablation of both the HFEB and the $\mathcal{L}_{freq}$ demonstrated that these components are critical for optimal prediction performance. Notably, the absence of the HFEB led to a more substantial degradation in performance compared to the removal of the AFFB, with a 0.510\% increase in MAE and a 0.904\% increase in RMSE. This underscores the pivotal role of the HFEB in capturing high-frequency details essential for accurate predictions, contributing more significantly to the model's overall performance than the AFFB.

Additionally, we explored the temporal and spatial distributions of MAE differences among the three comparative models. Fig. \ref{fig_FCNet_wo_AFFB} illustrates the spatial distribution of MAE differences between FCNet w/o AFFB and FCNet. Fig. \ref{fig_FCNet_wo_HFEB} presents the spatial distribution of MAE differences between FCNet w/o HFEB and FCNet. The MAE for each grid is averaged over all dates within the year. Red grids indicate that the MAE of the comparative model in that grid is greater than that of FCNet. Consequently, for most regions of the Arctic, both HFEB and AFFB have proven effective, with HFEB providing a greater benefit than AFFB.

\subsection{Performance of Recursive Prediction}

FCNet can recursively predict longer SIC sequences. We explore the performance of FCNet for recursive prediction. the predicted results for the first 14 days are used as inputs to predict the SIC for the following 14 days. Likewise, the predicted SIC for the next 14 days can be used as inputs to generate a 42-day SIC prediction. In this manner, we recursively predict longer SIC sequences.

We conducted extensive experiments for recursive prediction, and the experimental results are shown in Table \ref{table_recur}. As the recursive steps increase, the MAE and RMSE values increase, while NSE values decrease accordingly. It is evident that as the prediction time increases, the model's performance in predicting SIC worsens. However, for the 56-day SIC prediction, the maximum value of MAE is 4.207\%, which is still satisfying and relatively small. Therefore, we can draw the conclusion that the our FCNet also has strong capabilities for monthly SIC prediction.

To further evaluate the performance of the model in recursive prediction, we calculated the sea ice edge based on the predicted SIC and used the BACC metric \cite{icenet} to compute the proximity of the predicted ice edges and actual ones. BACC originates from the Integrated Ice Edge Error (IIEE). To obtain sea ice edges, grid cells with SIC values greater than 15\% are considered as sea ice, and grid cells with SIC values less than 15\% are considered as open water. Therefore, given the binary sea ice area judgment of SIC greater than 15\% (value 1) or not greater than 15\% (value 0), IIEE is defined as the sum of all areas where local sea ice extent is overestimated or underestimated \cite{goessling2016predictability}. BACC is a normalized version of IIEE, defined as:
\begin{equation}
    \text{BACC} = \left(1 - \frac{\text{IIEE}}{\text{area of active grid cell region}}\right) \times 100\%
\end{equation}
where the active grid cell area is the maximum daily sea ice extent (SIC greater than 15\%) observed from 1991 to 2020. We selected a 10-week period from August to October (70 days) to evaluate the model's performance. The experimental results are shown in Fig. \ref{fig_recur_metric}. For the first 14-day prediction, the average MAE values are less than 2\%. The BACC values decrease with the increase of recursive steps. Users can choose the appropriate prediction period based on their own needs.

Fig. \ref{fig_sie} shows the predicted sea ice extent of our FCNet for 2019, specifically focusing on four recursive predictions (SIC greater than 15\%). We compare the sea ice extent with PredRNN, IceNet and SEAS5. It can be observed that the sea ice extent predicted by our FCNet is the most closest to the ground truth.

\begin{table}[htbp]
    \centering
    \caption{Performance of minimum sea ice extent in three extreme cases ( Sep 2007, Sep 2012 and Sep 2020).}
    \scalebox{0.9}{
        \begin{tabular}{cccc}
            \multicolumn{4}{c}{MAE $\downarrow $}                                    \\
            \hline\toprule
            Method       & 2007              & 2012              & 2020              \\
            \midrule
            PredRNN      & 3.124\%           & 3.203\%           & 3.577\%           \\
            IceNet       & 2.901\%           & 2.986\%           & 3.240\%           \\
            FCNet (Ours) & \textbf{2.282\%}  & \textbf{2.326\% } & \textbf{2.481\% } \\
            \bottomrule\hline

            \multicolumn{4}{c}{ }                                                    \\

            \multicolumn{4}{c}{BACC $\uparrow $}                                     \\
            \hline\toprule
            Method       & 2007              & 2012              & 2020              \\
            \midrule
            PredRNN      & 91.103\%          & 90.631\%          & 89.278\%          \\
            IceNet       & 93.839\%          & 92.778\%          & 90.855\%          \\
            FCNet (Ours) & \textbf{96.826\%} & \textbf{96.685\%} & \textbf{95.395\%} \\
            \bottomrule\hline

            \multicolumn{4}{c}{ }                                                    \\
            \multicolumn{4}{c}{SIE(million km²)}                                     \\
            \hline\toprule
            Method       & 2007              & 2012              & 2020              \\
            \midrule
            Observation  & 4.170             & 3.387             & 3.262             \\
            \midrule
            PredRNN      & 4.723             & 3.956             & 3.907             \\
            IceNet       & 4.531             & 3.787             & 3.712             \\
            FCNet (Ours) & \textbf{4.328}    & \textbf{3.577}    & \textbf{3.471}    \\
            \bottomrule\hline
        \end{tabular}}
    \label{table_extremecase}
\end{table}

\subsection{Performance of FCNet in Extreme Cases}

We analyze the performance of our FCNet in predicting the minimum sea ice extent in three extreme cases (September 2007, September 2012, and September 2020). The minimum Arctic sea ice extent in September 2012 was unprecedented in the historical record. Furthermore, the Arctic sea ice extents in September 2007 and 2020 were also exceptionally low.

As shown in Fig. \ref{fig_extremecase}, we have analyzed in detail the sea ice coverage in the three extreme cases. Specifically, the minimum sea ice coverage observed in 2007 was 4.170 million square kilometers, and this extreme value occurred on September 18th. Our FCNet model significantly outperforms both PredRNN, SEAS5, and IceNet in this key metric. In addition, FCNet also shows excellent performance in two other extreme cases (i.e., September 17, 2012 and September 16, 2020).

According to the data in Table \ref{table_extremecase}, for the 14-day prediction task, FCNet achieves the best performance in all three extreme cases. Further, we delve into FCNet's prediction error, i.e., sea ice area error (SIEE), for these three extreme dates. In 2007, for example, the observed minimum sea ice area is 4.170 million square kilometers, and the prediction error of FCNet is only 0.158 million square kilometers, compared to 0.361 million square kilometers for IceNet, 0.457 million square kilometers for SEAS5, and 0.553 million square kilometers for PredRNN. In the other two extreme dates, FCNet also maintains a low error level: on September 17, 2012, the SIEE of FCNet is 0.21 million square kilometers lower than that of SEAS5 and 0.15 million square kilometers lower than that of IceNet; on September 16, 2020, the SIEE of FCNet is 0.241 million square kilometers lower than that of SEAS5 and 0.18 million square kilometers lower than that of IceNet. Therefore, it is confirmed that the proposed FCNet exhibits good performance in extreme minimum sea ice extent events, consistently outperforming SEAS5, PredRNN, and IceNet.

\section{Conclusion and Future Work}

In this paper, we propose the FCNet, a frequency-compensated network for daily sea ice concentration prediction across the Arctic region for a two-week period. Existing methods rarely explore the use of frequency domain features in this scenario, and thus have limited capability in extracting high-frequency details. To solve both problems, we design a dual-branch network, including two blocks for frequency feature extraction and convolutional feature extraction. For the frequency feature extraction, we design AFFB, which integrates trainable neural networks with Fourier-based filters. For the convolutional feature extraction, we propose HFEB to enhance the high-frequency feature information. Our FCNet was employed on a satellite-derived daily SIC data, and demonstrates superior performance in SIC prediction compared to the state-of-the-art models. In addition, the proposed FCNet exhibits exceptional performance in recursive prediction and extreme minimum sea ice extent cases.

It is noteworthy that remote sensing based SIC products are prone to be influenced by some uncertainty factors. However, the evaluation of SIC product uncertainty requires the use of multi-source remote sensing observations, which is not within the scope of this study. In the future, we plan to design SIC prediction models using multi-source remote sensing data. In addition, we plan to integrate physical knowledge of the ocean into the model to improve the interpretability.

\bibliography{source}
\bibliographystyle{IEEEtran}

\end{document}